\documentclass[aps,pra,10pt,superscriptaddress,notitlepage,nofootinbib,floatfix,twocolumn]{revtex4-2}

\usepackage{amsmath}
\usepackage{graphicx}
\usepackage{color}
\usepackage[FIGTOPCAP,raggedright,nooneline]{subfigure}
\usepackage{amsfonts}
\usepackage{dsfont}
\usepackage[normalem]{ulem}
\usepackage{multirow}
\newcommand{\ddx}[2][]{\frac{\partial^2 {#1}}{\partial #2^2}}
\newcommand{\abs}[1]{\left| {#1} \right|}

\newcommand{\E}{\mathcal{E}}

\newcommand{\dint}{\mathrm{d}}
\newcommand{\e}{\mathrm{e}}

\newcommand{\G}{\mathrm{G}}
\renewcommand{\L}{\mathrm{L}}
\newcommand{\ii}{\mathrm{i}}

\begin{document}
\title{Universality of excited three-body bound states in one dimension}

\author{Lucas Happ}
\email{lucas.happ@uni-ulm.de}
\affiliation{Institut f\"{u}r Quantenphysik and Center for Integrated Quantum Science and Technology ({$\rm IQ$}$^{\rm ST}$),  Universit\"{a}t Ulm, D-89069 Ulm, Germany}
\author{Matthias Zimmermann}
\affiliation{Institut f\"{u}r Quantenphysik and Center for Integrated Quantum Science and Technology ({$\rm IQ$}$^{\rm ST}$),  Universit\"{a}t Ulm, D-89069 Ulm, Germany}
\affiliation{Institute of Quantum Technologies, German Aerospace Center (DLR), D-89069 Ulm, Germany}
\author{Maxim A. Efremov}
\affiliation{Institut f\"{u}r Quantenphysik and Center for Integrated Quantum Science and Technology ({$\rm IQ$}$^{\rm ST}$),  Universit\"{a}t Ulm, D-89069 Ulm, Germany}
\affiliation{Institute of Quantum Technologies, German Aerospace Center (DLR), D-89069 Ulm, Germany}

\date{\today}

\begin{abstract}
We study a heavy-heavy-light three-body system confined to one space dimension provided the binding energy of an excited state in the heavy-light subsystems approaches zero. The associated two-body system is characterized by (i) the structure of the weakly-bound excited heavy-light state and (ii) the presence of deeply-bound heavy-light states. The consequences of these aspects for the behavior of the three-body system are analyzed. We find strong indication for universal behavior of both three-body binding energies and wave functions for different weakly-bound excited states in the heavy-light subsystems.
\end{abstract}

\maketitle

\section{Introduction}
Already relatively simple few-body systems, like a configuration of three pairwise interacting identical bosons in three dimensions, display rich features as the Efimov effect \cite{Efimov1970,Efimov1971}. This is the emergence of an infinite series of universal three-body bound states provided there are at least two $s$-wave resonant pair-interactions. Here, universality means that the three-body states are independent of the details of the two-body interaction \cite{Naidon2017,Braaten2006}. The Efimov effect  is enhanced in a two-component three-body system with a large mass ratio between the two species \cite{Efimov1973,Pires2014,Tung2014}. The existence of these Efimov states is strongly restricted to particular values of the total angular momentum \cite{Endo2011,Kartavtsev2007}, the dimension of space \cite{Nielsen2001,Pricoupenko2010,Rosa2018}, and the symmetry of the weakly-bound (or virtual) two-body state \cite{Nishida2013,Efremov2013,Zhu2013}, including combinations thereof. Changing any of these properties of the system may prohibit the Efimov effect, however there can still be universal three-body bound states \cite{Kartavtsev2007,Kartavtsev2008,Pricoupenko2010,Efremov2013}.

Recently, we have proven universality \cite{Happ2019} in a heavy-heavy-light three-body system, which is confined to one dimension (1D). The universality requires the heavy-light interactions to be tuned towards the ground-state threshold, that is the binding energy of the ground state approaches zero. Indeed, in this limit the three-body binding energies and wave functions for arbitrary short-range two-body interactions converge  to the respective ones found for the zero-range interaction.

In this article we study the universality of the same three-body system provided the heavy-light interactions are tuned towards an excited-state threshold, that is the binding energy of an excited state approaches zero. This situation occurs in most experiments employing ultracold atoms. In comparison to the case of the ground-state threshold this implies (i) that the weakly-bound heavy-light state can either be symmetric or antisymmetric and (ii) the presence of deeply-bound heavy-light states. We analyze such three-body systems by solving the Faddeev equations numerically for different finite-range potentials. For the three-body binding energies we identify the same universal behavior, irrespectively of whether the threshold is characterized by a symmetric or antisymmetric weakly-bound heavy-light state. In contrast, we demonstrate that the universal limit of the corresponding three-body wave functions is distinct for these two different symmetries. Moreover, within the separable approximation \cite{Weinberg1963,Sitenko1991,Mestrom2019} of the finite-range interactions, we find that the deeply-bound heavy-light states play a crucial role for the universal limits. The three-body bound states presented in this article belong to the class of so-called bound states in a continuum \cite{vonNeumann1929,Stillinger1975,Hsu2016}.

Our research supports the increasing interest in few- and many-body systems confined to low dimensions. Already many decades ago, the Tonks-Girardeau gas \cite{Tonks1936,Girardeau1960} and the Lieb-Liniger model \cite{Lieb1963,McGuire1964} have been considered, which are based on the zero-range contact interaction \cite{Girardeau2004}. Three-body systems in 1D have recently \cite{Mora2005,Moroz2015,Sowinski2019} attracted increasing attention. In particular, studies on the inclusion of three-body interactions \cite{Mazets2008,Guijarro2018,Nishida2018}, on three-body systems of two different species \cite{Kartavtsev2009,Happ2019}, and on the accuracy of adiabatic methods \cite{Mehta2014,Happ2019} have been performed.

Not only theoretical studies, but also experimental measurements of few-body effects in low dimensions are nowadays feasible. For this purpose, ultracold gases are confined via anisotropic traps leading to cigar- or tube-shaped configurations \cite{Bloch2008,Blume2012}. Moreover, the strength of interactions in ultracold gases can be tuned over the scope of many orders of magnitude with the help of Feshbach-resonances \cite{Chin2010} or so-called confinement-induced resonances \cite{Olshanii1998,Dunjko2011}. Experimental observations of few-body effects have been brought to an entire new level of detail by manipulating ultracold atoms on the single-atom scale \cite{Serwane2011,Reynolds2020} ensuring pure few-body effects.

Most theoretical studies \cite{Dodd1972,Mehta2007,Kartavtsev2009,Mehta2014,Nishida2018,Guijarro2018,Happ2019} of three-body systems confined along two directions are based on 1D models. This reduction of complexity offers the advantage of a simple and intuitive description revealing the underlying three-body properties. However, it is important to keep in mind that experiments employing these confined systems are always performed in quasi-1D. The conditions to reproduce the results predicted by a pure 2D model in a quasi-2D experiment and accordingly by a 1D model in a quasi-1D setup have been discussed in Refs. \cite{Levinsen2014,Blume2014,Sandoval2018}.

Our article is organized as follows. In section \ref{sec:statement} we introduce the class of three-body systems which is at the center of our study and precisely formulate our goals. We present then in section \ref{sec:Results} the corresponding energies and wave functions of three-body bound states in the system. In particular, we analyze the universal behavior of these quantities close to different energy thresholds of the two-body system. Next, we discuss in section \ref{sec:deeplybound} the influence of deeply bound two-body states on the universal behavior of the three-body system. Finally, we conclude by summarizing our results and by presenting an outlook in section \ref{sec:conclusion}. In order to keep our article self-contained, but focused on the central ideas we collect in Appendix \ref{app:formalism} the methods used to solve the three-body system.

\section{The three-body system}\label{sec:statement}
In this section we first introduce a one-dimensional two-body system governed by a pair interaction. Next, by adding a third particle we arrive at the class of three-body systems which is at the focus of this article. We then present characteristic quantities of the three-body system which are central to our subsequent studies. Finally, we describe the analytical and numerical methods employed for the determination of the three-body binding energies and wave functions. 

\subsection{Two interacting particles}\label{sec:twobody1D}
We consider a two-body system consisting of a heavy particle of mass~$M$ and a light one of mass~$m$, both constrained to 1D and interacting via a potential of finite range $\xi_0$. We define the mass ratio $\alpha \equiv M/m$ of the two particles in the heavy-light system.

After eliminating the heavy-light center-of-mass coordinate, the system is governed by the stationary Schr\"odinger equation 
\begin{equation}
\left[-\frac{1}{2}\frac{\textrm{d}^2}{\textrm{d}\xi^2}+v\left(\xi \right)\right]\psi^{(2)}=\mathcal{E}^{(2)}\psi^{(2)}
\label{eq:Schroedinger_2body_dmless}
\end{equation}
for the two-body wave function $\psi^{(2)}= \psi^{(2)}(\xi)$ of the relative motion presented in dimensionless units. Here, $\xi$ denotes the relative coordinate of the light particle with respect to the heavy one in units of the characteristic length $\xi_0$.

The two-body binding energy $\mathcal{E}^{(2)}$ and the  potential 
\begin{equation}\label{interaction}
v\left(\xi \right)= v_0 f(\xi)
\end{equation}
are both given in units of $ \hbar^2/\mu \xi_0^2$ with the Planck constant $\hbar$ and the reduced mass $\mu\equiv M/(1+\alpha)$ of the heavy-light system. Here, $v_0$ denotes the magnitude and $f$ the shape of the interaction potential.

We assume an attractive interaction with $v_0 < 0$ and $\int \dint \xi f(\xi) > 0$, as well as a symmetric shape $f$, that is $f(\xi) = f(\abs{\xi})$. Moreover, we choose $v$ such that it describes a short-range interaction, \textit{i.e.} $\abs{\xi}^2 f(|\xi|) \to 0$ as $\abs{\xi} \to \infty$.

In particular, we perform the analysis in this article for different two-body interactions of finite-range, namely a potential having the polynomially decaying shape
\begin{equation}\label{eq:fL}
f_\mathrm{L}(\xi) \equiv \frac{1}{(1+\xi^2)^3}
\end{equation}
of the cube of a Lorentzian, and one with the shape
\begin{equation}\label{eq:fG}
f_\mathrm{G}(\xi) \equiv \exp(-\xi^2)
\end{equation}
of a Gaussian decaying exponentially.

In contrast to the zero-range interaction of shape
\begin{equation}\label{eq:fD}
f_\delta(\xi) \equiv \delta(\xi),
\end{equation}
the finite-range potentials whose shapes are determined by Eqs. \eqref{eq:fL} and \eqref{eq:fG} can support more than a single bound state, depending on the magnitude $v_0$.

We label the two-body wave function $\psi_r^{(2)}$ and the binding energy $\E_r^{(2)}$ as solutions of Eq. \eqref{eq:Schroedinger_2body_dmless} by the index $r$, where $r=0,1,2,\ldots$ denotes the number of nodes of $\psi_r^{(2)}$. Even values of $r$ indicate symmetric two-body wave functions, whereas odd values indicate antisymmetric ones, that is 
\begin{equation}\label{eq:evenodd}
\psi_r^{(2)}(-\xi) = (-1)^r \psi_r^{(2)}(\xi).
\end{equation}

\subsection{Three interacting particles}
\label{sec:threebody1D}
\begin{figure}[htbp]
\includegraphics[width =.8\columnwidth]{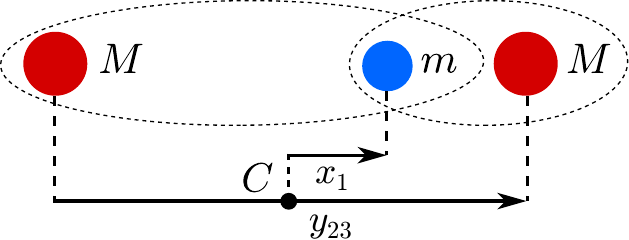}
\caption{Jacobi-coordinates $x_1$ and $y_{23}$ for the three-body system consisting of one light particle of mass $m$ (blue) and two heavy ones of mass $M$ (red), all confined to 1D. Here, $C$ denotes the center of mass of the two heavy particles.}
\label{fig:coords}
\end{figure}

We now add a third particle to the two-body system, also constrained to 1D and identical to the other heavy particle of mass $M$. We assume the same interaction between each heavy and the light particle as introduced in section \ref{sec:twobody1D}, but no interaction between the two heavy ones. The resulting three-body system is depicted in Fig. \ref{fig:coords}. Here, $y_{23}$ denotes the relative coordinate between the two heavy particles (called here particles 2 and 3) and $x_1$ the relative coordinate of the light particle (particle 1) with respect to the center of mass $C$ of the two heavy ones, both in units of the characteristic length $\xi_0$.

Eliminating the center-of-mass motion of this heavy-heavy-light system, we arrive at the Schr\"odinger equation
\begin{align}\label{eq:3bodySGL}
\left[-\frac{\alpha_x}{2}\ddx{x_1} - \frac{\alpha_y}{2}\ddx{y_{23}} + v\left(x_1+\frac{y_{23}}{2}\right) + v\left(x_1-\frac{y_{23}}{2}\right)\right]\psi& \nonumber\\
 = \E \psi&
\end{align}
for the three-body wave function $\psi = \psi(y_{23},x_1)$ of the two relative motions in coordinate representation. The coefficients $\alpha_x \equiv (1+2\alpha)/[2(1+\alpha)]$ and $\alpha_y \equiv 2/(1+\alpha)$ depend only on the mass ratio $\alpha$, and $\E$ is the dimensionless three-body energy in units of $ \hbar^2/\mu \xi_0^2$.

\subsection{Formulation of the problem}
In this article we determine the energies and wave functions of bound states in the three-body system, provided the heavy-light subsystems support a weakly bound excited state, that is $\E_r^{(2)} \to 0^-$ for $r=1,2,3$. We are mainly interested in the question if in these cases there exist three-body quantities that show universal behavior. Universality denotes the property of physical quantities to show a behavior that is independent of the short-range details of the interparticle interactions.

In Ref. \cite{Happ2019} we have studied the situation when the heavy-light system is tuned to the ground-state threshold ($r=0$). There we have found universality in terms of both, energies and corresponding wave functions of three-body bound states for $\E_0^{(2)}\to 0^-$. In this limit we have proven that for arbitrary attractive short-range heavy-light interactions obeying the conditions in section \ref{sec:twobody1D}, the three-body energies and wave functions converge to the respective ones induced by a heavy-light contact interaction of shape $f_\delta$, Eq. \eqref{eq:fD}.

The differences between the ground- and an excited-state threshold are as follows. Tuning the heavy-light subsystems to an excited-state energy threshold immediately implies (i) a different number $r$ of nodes and therefore a possibly different symmetry of the corresponding weakly-bound two-body bound state, and (ii) the existence of deeply bound two-body states, whose binding energy does not approach zero. Depending on the symmetry of the weakly-bound two-body wave function, Eq.~\eqref{eq:evenodd}, we speak of even-numbered ($r=0,2,\ldots$) and odd-numbered ($r=1,3,\ldots$) thresholds.

We analyze the effect of both features on the universal behavior of energies and wave functions of three-body bound states close to an excited-state threshold in the heavy-light subsystems. We emphasize that there are two kinds of three-body bound states, namely (i) excited three-body bound states associated with the respective threshold, and (ii) deeply-lying three-body bound states. (i) The excited three-body bound states are embedded in a continuum that originates from deeply bound heavy-light states in combination with an unbound heavy particle. For these states we expect a universal behavior. These so-called bound states in a continuum have already been predicted and experimentally realized for a variety of physical systems \cite{vonNeumann1929,Stillinger1975,Hsu2016}. (ii) The deeply bound three-body states keep a finite size when the excited-state threshold is approached, hence we expect them to be sensitive to the details of the interaction. For this reason they do not show universal behavior and we refrain from analyzing them in this article.

Although not being studied in the present article, we want to mention that there exists also the possibility of three-body resonances \cite{Csoto1994,Papp2005,Garrido2005} being located in the same energy regime as the aforementioned excited three-body bound states. These resonances are characterized by a non-vanishing imaginary part of the three-body energy.

Similar to the results \cite{Happ2019} found near the ground-state threshold ($\E_0^{(2)} \to 0^-$), we expect a set of universal three-body bound states associated to each heavy-light threshold ($\E_r^{(2)} \to 0^-$) characterized by the index $r$. To distinguish between these sets, we introduce the notation $\E_{r,n}$ for the energy of the $n$-th three-body bound state $|\psi_{r,n}\rangle$ within the $r$-th set. As shown in the following, we indicate with even and odd values of $n$ the case of the two heavy particles being bosons or fermions, respectively.

In order to reveal the dependency of three-body universality on (i) the symmetry of the weakly-bound state and (ii) the existence of deeply bound states in the heavy-light subsystems, we compare the energies and wave functions of three-body bound states for $r>0$ with the corresponding ones for $r=0$. Due to the reported \cite{Happ2019} universality for $r=0$, we can use the scale-invariant three-body binding energies and wave functions for the heavy-light contact interaction as representatives for this set of states. We therefore base our analysis on the following two quantities:

(a) The relative deviation
\begin{equation}\label{eq:relativedeviation}
\Delta\epsilon_{r,n} \equiv \abs{\frac{\epsilon_{r,n} - \epsilon_n^\star}{\epsilon_n^\star}}
\end{equation}
between the  energy ratios $\epsilon_{r,n}$ and $\epsilon_n^\star$. Here,
\begin{equation}\label{epsilon}
\epsilon_{r,n}\equiv \frac{\mathcal{E}_{r,n}}{\left|\mathcal{E}_{r}^{(2)}\right|}
\end{equation}
denotes the ratio of the $n$-th three-body bound state energy $\E_{r,n}$ within the $r$-th set to the energy $\E_{r}^{(2)}$ of the weakly-bound heavy-light bound state. Moreover, we define the ratio
\begin{equation}\label{eq:epsilon_star}
\epsilon_n^\star \equiv \frac{\mathcal{E}_{0,n}}{\left|\mathcal{E}_{0}^{(2)}\right|}
\end{equation}
between the corresponding three- and two-body energies found for the special case of a contact heavy-light interaction of shape $f_\delta$, Eq. \eqref{eq:fD}.

(b) The fidelity
\begin{equation}\label{eq:fidelity}
F_{r,n} \equiv \abs{\langle \psi_{r,n}|\psi_n^\star\rangle}^2
\end{equation}
defined as the overlap between a three-body bound-state $|\psi_{r,n}\rangle$ for a finite-range heavy-light potential, and the corresponding state $|\psi_{n}^\star \rangle$ for the contact interaction.

\subsection{Methods}\label{sec:methods}

We solve Eq. \eqref{eq:3bodySGL} within the framework of the Faddeev equations \cite{Faddeev1961,Sitenko1991}. There we can easily incorporate the necessary boundary condition 
\begin{equation}\label{eq:BoundaryCondition}
\psi(y_{23},x_1) \to 0\, ,\qquad \mathrm{as} \, \abs{x_1} \to \infty\,\  \mathrm{and} \ \abs{y_{23}} \to \infty,
\end{equation}
to obtain three-body bound states embedded in a continuum. For this reason we consider the homogeneous form of the Faddeev equations.

We show in Appendix \ref{app:formalism} that the wave function $\psi(k_{23},p_1)$ of a three-body bound state in momentum representation is given by the superposition
\begin{align}\label{eq:superposition_MainText}
\psi(k_{23},p_1) =~&\phi^{(2)}\left(-\frac{\alpha_y}{2}k_{23} - \alpha_x p_1,k_{23}-\frac{1}{2}p_1\right) \nonumber \\
\pm& \phi^{(2)}\left(\frac{\alpha_y}{2}k_{23} - \alpha_x p_1,-k_{23}-\frac{1}{2}p_1\right)
\end{align}
of the Faddeev component $\phi^{(2)}$ evaluated at different arguments. Here, $k_{23}$ denotes the relative momentum between particles 2 and 3, whereas $p_{1}$ is the momentum of particle 1 relative to the center of mass of particles 2 and 3.

Moreover, the plus and minus signs in Eq. \eqref{eq:superposition_MainText} distinguish the case of the two heavy particles being bosons or fermions. This is evident from considering the exchange of the two heavy particles being described by $k_{23} \to -k_{23}$ which leads to the exchange symmetry
\begin{equation}\label{eq:bosferm_MainText}
\psi(-k_{23},p_1) = \pm \psi(k_{23},p_1)
\end{equation}
for the total three-body wave function. As $k_{23}$ is the momentum corresponding to the relative distance $y_{23}$ between the two heavy particles, the symmetry (antisymmetry) of $\psi(k_{23},p_{1})$ with respect to the line $k_{23}=0$ directly relates via the Fourier transform to the symmetry (antisymmetry) of the wave function $\psi(y_{23},x_{1})$ with respect to the line $y_{23}=0$ in coordinate space.

The separable expansion \cite{Sitenko1991,Weinberg1963,Mestrom2019} simplifies the Faddeev equations (see Appendix \ref{app:separableexpansion}) and allows to analyze the influence of deeply bound two-body states on three-body universality. We therefore expand the Faddeev component
\begin{equation}\label{eq:sep_phi_MainText}
\phi^{(2)}(k,p) \equiv \pm \frac{1}{\E_p - \frac{1}{2} k^2}\sum_{\nu=0}^{\infty} g_\nu(k,\E_p)\tau_\nu(\E_p)\varphi_\nu(p,\E)
\end{equation}
in terms of products of functions depending on only one momentum variable. Here, $\E_p\equiv \E - \alpha_x \alpha_y\,p^2/2$ and the functions $g_\nu$ and $\tau_\nu \equiv -\eta_\nu/(1-\eta_\nu)$ are determined by the eigenvalue equation
\begin{equation}\label{eq:etaeigenvalueeq_MainText}
\int \frac{\dint k'}{2\pi} V(k,k') \frac{1}{\E_p-\frac{1}{2}k'^2}\,g_\nu(k',\E_p) = \eta_\nu(\E_p) g_\nu(k,\E_p),
\end{equation}
where
\begin{equation}\label{eq:PotentialInMomentum_MainText}
V(k,k') = v_0 \int \dint \xi \, f(\xi) \e^{-\ii(k-k')\xi}
\end{equation}
is the momentum representation of the heavy-light potential $v(\xi)$, Eq. \eqref{interaction}.

For a given $r>0$ and $\nu \leq r$ the functions $g_\nu$ can be related to the wave functions $\psi_\nu^{(2)}$ of the bound two-body states, however with variable energy argument $\E_q$. Indeed, for a given $v_0$, they are connected to each other only when $\E_q = \E_\nu^{(2)}$ \cite{Weinberg1963,Sitenko1991}. The term with $\nu=r$ describes the one close to the threshold and the terms with $\nu < r$ correspond to the deeply bound heavy-light states. We can reveal the influence of the latter for the prediction of the correct three-body universality by excluding single terms with $\nu < r$ in the expansion, Eq. \eqref{eq:sep_phi_MainText}, and comparing the subsequent results to the ones without excluding them. This is performed in section \ref{sec:deeplybound}.

For the functions $\varphi_\nu$ we derive in Appendix \ref{app:separableexpansion} the system of coupled integral equations
\begin{align}\label{eq:FE_final_MainText}
\varphi_\lambda(p,\E) = \pm &\sum_{\nu=0}^{\infty}\int \frac{\dint q}{2\pi}\, \tau_{\nu}(\E_{q})\, \varphi_{\nu}(q,\E)  \nonumber \\
~&\times \frac{g_{\lambda}\left(q+\frac{\alpha}{1+\alpha}p,\E_p\right)\, g_{\nu}\left(p + \frac{\alpha}{1+\alpha}q,\E_{q}\right)}{\E-\frac{1}{2}q^2 - \frac{1}{2}p^2 - \frac{\alpha}{1+\alpha}p q}
\end{align}
which yields the same energy solutions $\E_{r,n}$ as the Schr\"odinger equation \eqref{eq:3bodySGL}.

As mentioned above, we consider in this article three-body bound states with the boundary conditions, Eq. \eqref{eq:BoundaryCondition}, embedded in a continuum associated to deeply bound heavy-light states together with an unbound heavy particle. They therefore correspond to real energy solutions $\E_{r,n}$. The three-body bound state energy $\E$ enters the kernels in Eq. \eqref{eq:FE_final_MainText} as a parameter, hence the solutions $\E_{r,n}$ are obtained by varying over a suitable region $\E_{r,n} < \E_r^{(2)}$ and requiring eigenvalue unity (minus unity) for bosons (fermions). The functions $\varphi_\nu$ are obtained as the corresponding eigenvectors.

Equation \eqref{eq:FE_final_MainText} is a system of coupled homogeneous Fredholm integral equations of the second kind. We solve it numerically by truncating the infinite number of expansion terms in Eqs. \eqref{eq:sep_phi_MainText} and \eqref{eq:FE_final_MainText} to a finite value $\nu_\mathrm{max}$, and approximating the continuous interval for $p$ and $q$ by a discrete grid. This reduces Eq. \eqref{eq:FE_final_MainText} to a finite system of coupled matrix eigenvalue problems. In our analysis $\nu_\mathrm{max} = 10$ has yielded sufficient convergence.

\section{Results}\label{sec:Results}
In this section we present the energies and wave functions of three-body bound states associated to the two-body threshold $\E_r^{(2)} \to 0^-$ for  $r=1$, $r=2$ and $r=3$. For this purpose, we have solved Eq.~\eqref{eq:FE_final_MainText} numerically for two finite-range interaction potentials of shape $f_\mathrm{L}$,~Eq. \eqref{eq:fL}, and $f_\mathrm{G}$, Eq.~\eqref{eq:fG}. By varying the magnitude $v_0$ of the potential, we can tune the heavy-light subsystems to a particular threshold.

The mass ratio between heavy and light particles directly influences the number of three-body bound states \cite{Kartavtsev2009,Happ2019}. In the following we choose a mass ratio of $\alpha = M/m = 20$ for which there is a total of six three-body bound states associated to the ground-state heavy-light threshold, three each for the case of bosonic and fermionic heavy particles. In the next subsections we show that the number of respective three-body bound states in one set remains the same near an excited-state heavy-light threshold.

The ratios $\epsilon_n^\star$, Eq. \eqref{eq:epsilon_star}, obtained for a contact heavy-light interaction and the mass ratio $M/m = 20$ are listed in Table \ref{tab:epsilonstar}. Due to the scale-invariant nature of the delta-potential, they are independent of the heavy-light binding energy. Since all three-body binding energies associated with a weakly bound heavy-light state lie below the corresponding two-body threshold, all ratios are smaller than $-1$.

\begin{table}[htbp]
\caption{\label{tab:epsilonstar}Energy ratios $\epsilon_n^\star$, Eq. \eqref{eq:epsilon_star}, for the two-body contact interaction $f_\delta$, Eq. \eqref{eq:fD}, and the mass ratio $M/m = 20$. As shown in Ref. \cite{Happ2019}, they represent the universal limits obtained for attractive short-range heavy-light interactions in the case $\E_0^{(2)} \to 0^-$.}
\begin{ruledtabular}
\begin{tabular}{ccc}  
$n$    & Bosons & Fermions \\
\hline
%\cline{2-4}
\ ~0 \ \ & -2.7238	& -   \\
1 & -				& -1.6517  \\
2 & -1.3285 		& -  \\
3 & - 				& -1.1240  \\
4 & -1.0373 		& -  \\
5 & - 				& -1.0004 	   \\
\end{tabular}
\end{ruledtabular}
\end{table}

\subsection{Energy spectrum}\label{sec:EnergySpectrum}
\begin{figure*}[htbp]
\centering
\subfigure[\hspace{\columnwidth}]{
\includegraphics[width=.45\textwidth]{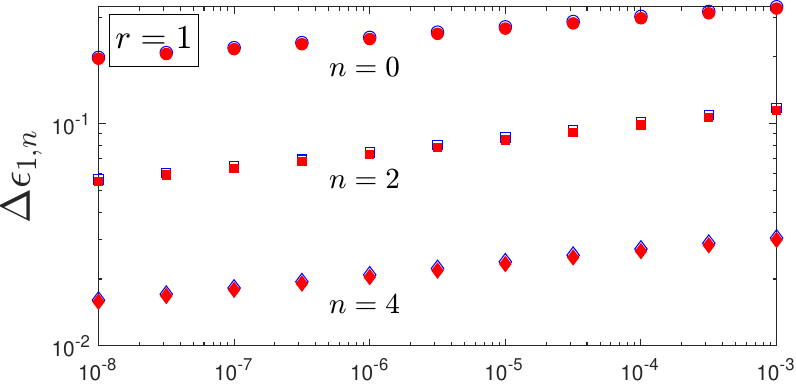}}
\hfill
\subfigure[\hspace{\columnwidth}]{
\includegraphics[width=.45\textwidth]{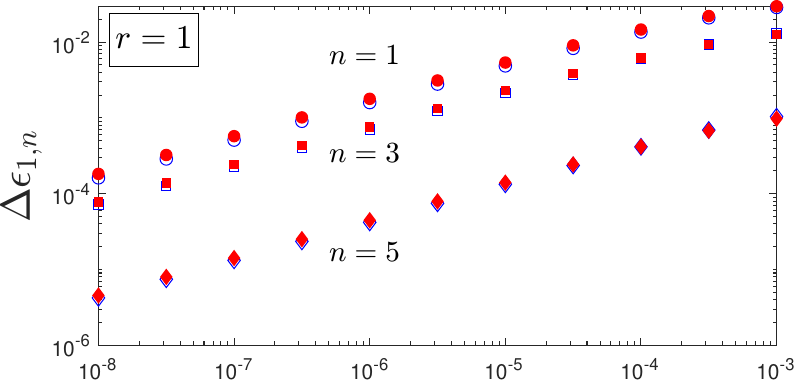}}\\
\subfigure[\hspace{\columnwidth}]{
\includegraphics[width=.45\textwidth]{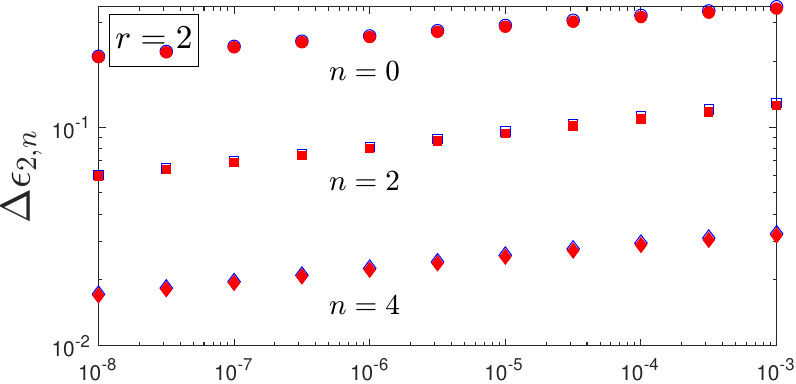}}
\hfill
\subfigure[\hspace{\columnwidth}]{
\includegraphics[width=.45\textwidth]{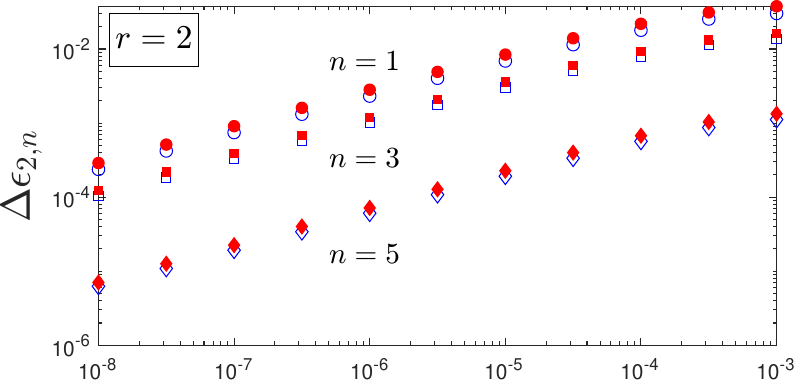}}\\
\subfigure[\hspace{\columnwidth}]{
\includegraphics[width=.45\textwidth]{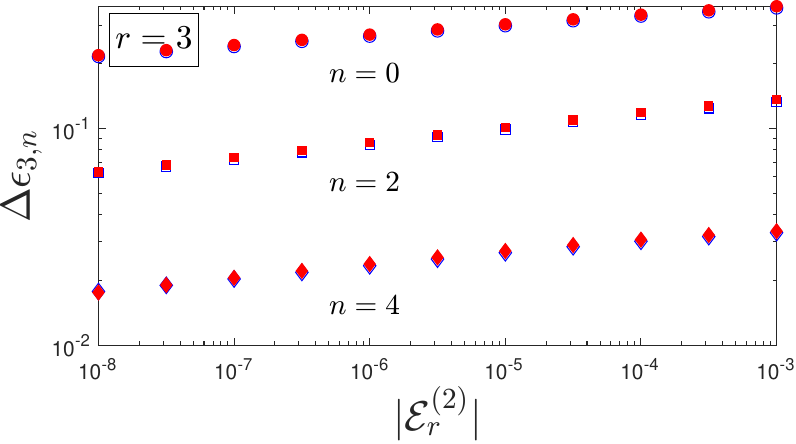}}
\hfill
\subfigure[\hspace{\columnwidth}]{
\includegraphics[width=.45\textwidth]{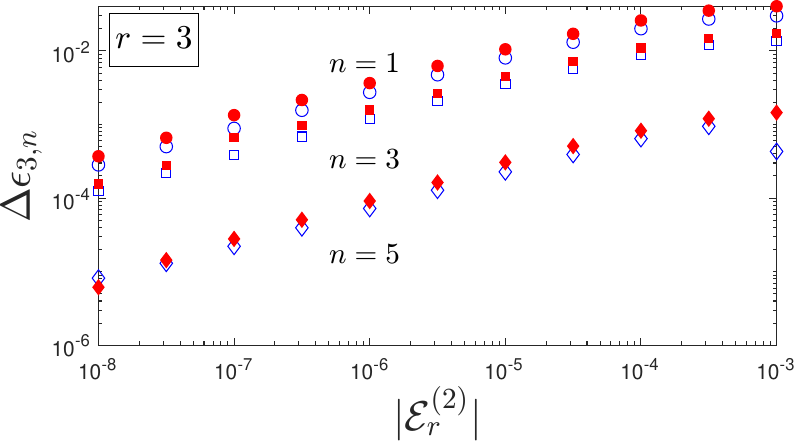}}
\caption{Relative deviation $\Delta \epsilon_{r,n}$, Eq. \eqref{eq:relativedeviation}, of the energy ratio $\epsilon_{r,n}$, Eq. \eqref{epsilon}, from the energy ratio $\epsilon_{n}^\star$, Eq. \eqref{eq:epsilon_star}, as a function of the two-body binding energy $\E_r^{(2)}$. Here we present the results for the excited-state thresholds with $r=1$ (top row),  $r=2$ (center row), and $r=3$, (bottom row). The left and right columns distinguish between the case of two heavy bosons and two heavy fermions. The results are shown for finite-range potentials of shape $f_\mathrm{L}$, Eq.~\eqref{eq:fL}, (filled red symbols) and $f_\mathrm{G}$, Eq.~\eqref{eq:fG}, (empty blue symbols). For a mass ratio of $M/m=20$ there are six three-body bound states associated to the individual heavy-light thresholds, three each for the case of bosons or fermions.% The results for the two finite-range potentials are almost identical. As $\E_r^{(2)} \to 0^-$, we see that in all cases the deviation $\Delta \epsilon_{r,n}$ decreases strictly monotonically. No significant difference is observed between the results for the three excited-state thresholds ($r=1,2,3$). For fermions, $\Delta \epsilon_{r,n}$ decreases much faster for $\E_r^{(2)}\to0^-$ in comparison to bosons, as indicated by the different range of the vertical axes.
}
\label{fig:spectrum_R23}
\end{figure*}

We start by analyzing the universality in the energy spectrum of the three-body system. In Fig. \ref{fig:spectrum_R23} we present the relative deviation $\Delta\epsilon_{r,n}$, Eq. \eqref{eq:relativedeviation}, between the energy ratio $\epsilon_{r,n}$, Eq. \eqref{epsilon}, obtained in the case of a finite-range heavy-light interaction and the energy ratio $\epsilon_n^\star$, Eq. \eqref{eq:epsilon_star}, for a zero-range heavy-light interaction, Eq. \eqref{eq:fD}, as a function of the  two-body binding energy $\E_r^{(2)}$. Throughout the subfigures we indicate by filled red and empty blue symbols the results for an interaction of shape $f_\L$, Eq. \eqref{eq:fL}, and $f_\mathrm{G}$, Eq. \eqref{eq:fG}, respectively. The top row shows $\Delta\epsilon_{1,n}$ near the first excited heavy-light threshold, $\E_1^{(2)}\to 0^-$, the central row $\Delta\epsilon_{2,n}$ for $\E_2^{(2)}\to0^-$, and the bottom row $\Delta\epsilon_{3,n}$ for the third excited state, $\E_3^{(2)}\to 0^-$. The left and right columns display the case of two heavy bosons ($n=0,2,4$) or fermions ($n=1,3,5$), respectively.

For all three thresholds ($r=1,2,3$) and all three-body bound states ($n=0\ldots 5$), the filled red and empty blue data sets are very close to each other. We highlight that this is true throughout the entire range of two-body energies displayed here. Moreover, the deviation between the red and blue data sets is decreasing as the threshold is approached. Thus, universality is indicated by the fact that near each threshold, the same energy ratios are achieved for the different finite-range interactions.

Next, we compare the universal behavior near the different heavy-light thresholds. For all considered cases $r=1,2,3$, the relative deviations $\Delta \epsilon_{r,n}$ are decreasing strictly monotonically towards zero as $\E_r^{(2)}\to0^-$. Hence, the energy ratios $\epsilon_{r,n}$ associated to all three excited-state two-body thresholds converge towards the same limit values $\epsilon_{n}^\star$, Table \ref{tab:epsilonstar}, as for the ground-state threshold. This explicitly implies that the universal limits of the three-body energies are identical for symmetric ($r=0,2,\ldots$) and antisymmetric ($r=1,3,\ldots$) weakly-bound states in the heavy-light subsystems. Moreover, we observe that for each three-body state labeled by $n$, the results for the considered thresholds $r=1,2,3$ behave almost identical throughout the entire range of two-body binding energies.

\begin{table}[htbp]
\caption{\label{tab:universalbehavior}Exponents of the suggested power law behavior of $\Delta\epsilon_{r,n}$, Eq. \eqref{eq:relativedeviation}, corresponding to the slopes of the linear dependence in the double-logarithmic scale, Fig. \ref{fig:spectrum_R23}. These values have been obtained via a linear regression for the potential of Gaussian shape $f_\G$, Eq. \eqref{eq:fG}. The corresponding results for $r=0$ are analyzed in more detail in Ref.~\cite{Happ2019} and not shown explicitly in Fig. \ref{fig:spectrum_R23}.}
%\begin{ruledtabular}
%\begin{tabular}{ccc}  
%$r$    & Bosons & Fermions \\
%\hline
%%\cline{2-4}
%\ ~0 \ \ 	& 1.026		& 0.989  \\
%1 			& 0.056		& 0.471  \\
%2 			& 0.057 	& 0.473  \\
%3 			& 0.056 	& 0.474  \\
%\end{tabular}
%\end{ruledtabular}
\begin{ruledtabular}
\begin{tabular}{cccccc}  
$r$    & $n$ & Bosons & \ ~ &$n$ & Fermions \\
\hline
%\cline{2-4}
\ ~\multirow{3}*{0} \ \	& 0	& 0.992	& &1 & 0.986  \\
						& 2	& 1.087	& &3 & 0.991  \\
						& 4	& 1.000	& &5 & 0.989  \\
\hline
\multirow{3}*{1}		& 0	& 0.046	& &1 & 0.461  \\
						& 2	& 0.065	& &3 & 0.463  \\
						& 4	& 0.057	& &5 & 0.488  \\
\hline
\multirow{3}*{2}		& 0	& 0.046	& &1 & 0.475  \\
						& 2	& 0.067	& &3 & 0.475  \\
						& 4	& 0.057	& &5 & 0.468  \\
\hline
\multirow{3}*{3}		& 0	& 0.046	& &1 & 0.469  \\
						& 2	& 0.067	& &3 & 0.469  \\
						& 4	& 0.056	& &5 & 0.485  \\	
\end{tabular}
\end{ruledtabular}
\end{table}

In particular, the deviations $\Delta \epsilon_{r,n}$ appear to decrease as a power law, as indicated by the linear dependence in the double-logarithmic scale. The corresponding slopes, which in return denote the exponent of the power law, are listed in Table \ref{tab:universalbehavior} and have been obtained via linear regression and for the potential of Gaussian shape $f_\G$, Eq.~\eqref{eq:fG}. The slopes for $r>0$ are smaller compared to those for $r=0$. This relates to a slower convergence towards the universal three-body energies near excited-state thresholds compared to the ground-state one. The difference between the thresholds $r=0$ and $r>0$ is characterized by the presence of deeply-bound states in the latter cases. Hence these deeply bound states have a crucial influence in how fast the universal limit is approached. Moreover, for each respective state labeled by $n$, we observe that the slopes are almost identical for all $r>0$. We note here, that $\Delta \epsilon_{r,n}$, Eq. \eqref{eq:relativedeviation}, is defined as an absolute value, hence our analysis does not distinguish whether the limit values $\epsilon_n^\star$ are approached from above or below. % The effect of these states on the limit value is discussed in more detail in section \ref{sec:deeplybound}.

We can further analyze the universal behavior for different excited three-body states characterized by the index $n$. In the case of fermions (odd $n$) the relative deviation of $\epsilon_{r,n}$ from $\epsilon_n^\star$ is much smaller and decreases much faster than in the case of bosons (even $n$)  when approaching the threshold, as indicated by the different range of the vertical axes. Indeed, this is also reflected in the values of the corresponding slopes, Table \ref{tab:universalbehavior}, which are almost a factor of ten larger for fermions compared to bosons. This striking difference in the universal behavior is a pure three-body effect due to the fact that the property of the two heavy particles being bosons or fermions does not change the heavy-light two-body problem. Finally, we observe a smaller deviation $\Delta\epsilon_{r,n}$ for states with larger index $n$.

\subsection{Wave functions}
Having found that the universal limit of the three-body binding energies is the same when even-numbered and odd-numbered thresholds are approached, we turn now to the analysis in terms of the corresponding three-body wave functions. Since the wave functions carry the full information about a quantum state, we aim to obtain a deeper insight into three-body universality from their study.

For this purpose, we compare the three-body wave functions for the two finite-range potentials of shape $f_\L$, Eq. \eqref{eq:fL}, and $f_\G$, Eq. \eqref{eq:fG}, tuned near the excited-state thresholds $r=1,2,3$, to the ones resulting from a contact heavy-light interaction with $f_\delta$, Eq. \eqref{eq:fD}, by means of the fidelity $F_{r,n}$ defined by Eq. (\ref{eq:fidelity}). As previously mentioned we use the results for the contact interaction as scale-invariant representatives for the case $r=0$. Moreover, in order to compare the universal behavior for weakly-bound excited heavy-light states of the same symmetry, we consider the overlap $\abs{\langle\psi_{3,n}|\psi_{1,n}\rangle}^2$ of the three-body wave functions for the two odd-numbered thresholds $r=1$ and $r=3$.

\begin{figure*}[htbp]
\centering
\subfigure[\hspace{\columnwidth}]{
\includegraphics[width=.45\textwidth]{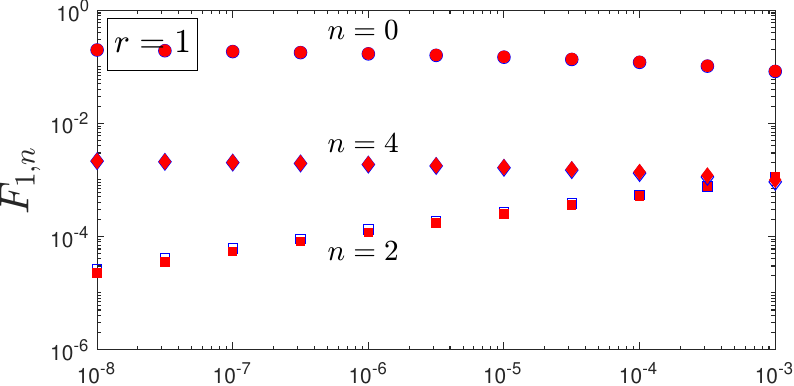}}
\hfill
\subfigure[\hspace{\columnwidth}]{
\includegraphics[width=.45\textwidth]{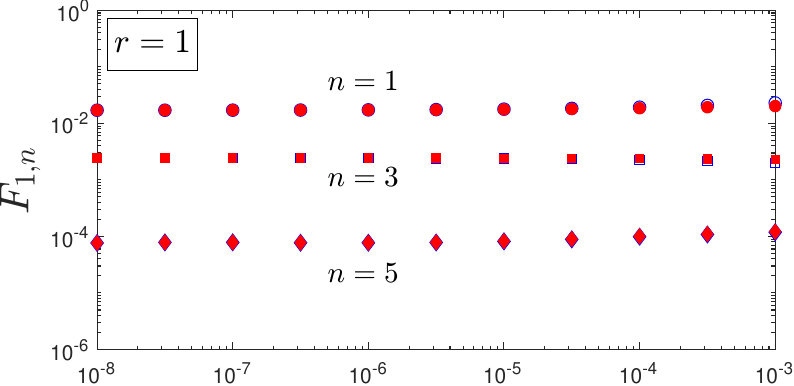}} \\
\subfigure[\hspace{\columnwidth}]{
\includegraphics[width=.45\textwidth]{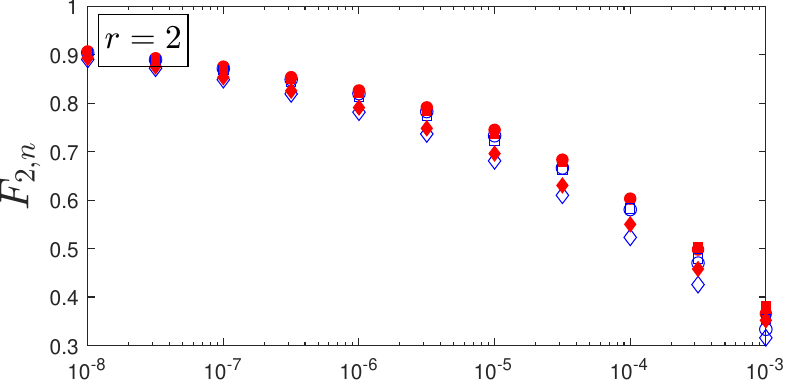}}
\hfill
\subfigure[\hspace{\columnwidth}]{
\includegraphics[width=.45\textwidth]{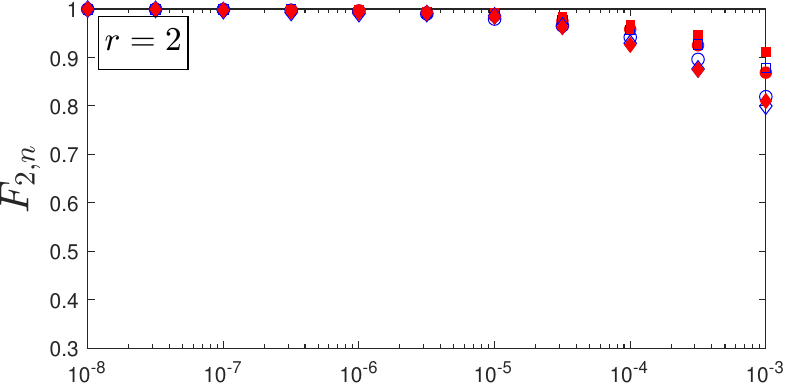}}\\
\subfigure[\hspace{\columnwidth}]{
\includegraphics[width=.45\textwidth]{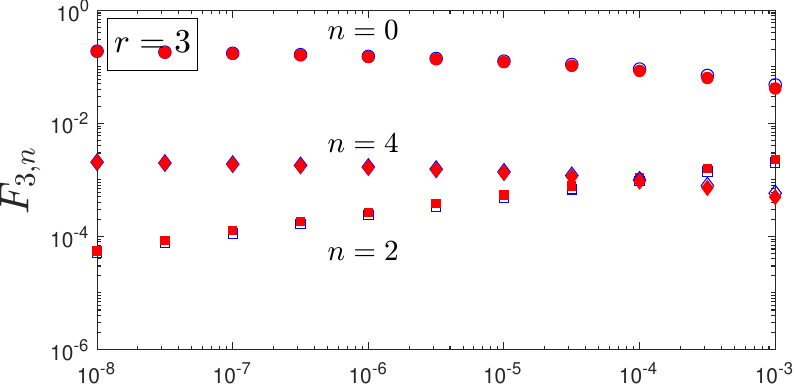}}
\hfill
\subfigure[\hspace{\columnwidth}]{
\includegraphics[width=.45\textwidth]{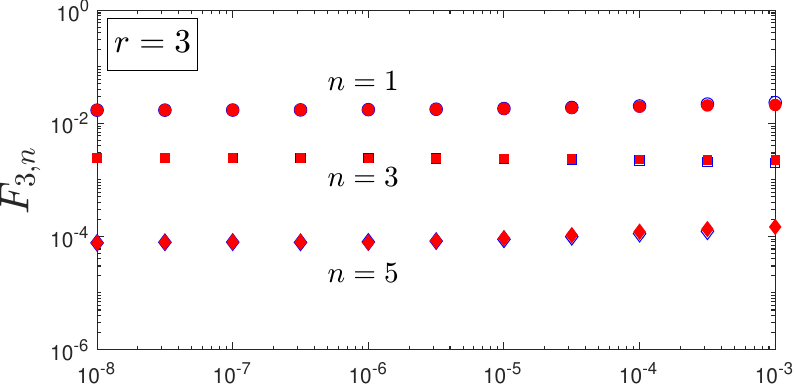}}\\
\subfigure[\hspace{\columnwidth}]{
\includegraphics[width=.45\textwidth]{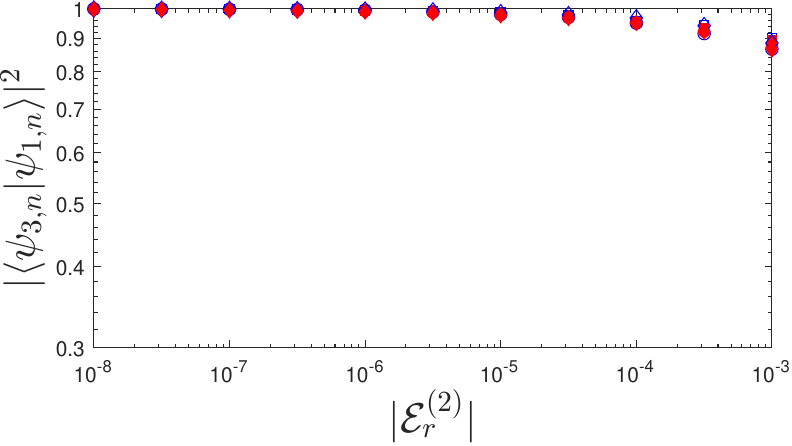}}
\hfill
\subfigure[\hspace{\columnwidth}]{
\includegraphics[width=.45\textwidth]{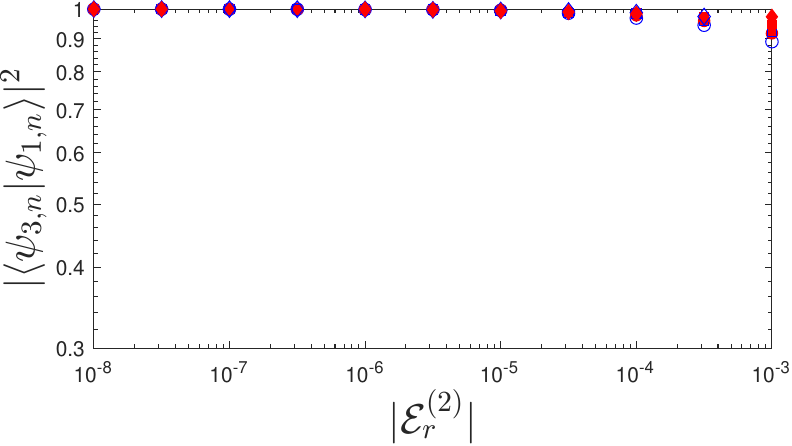}}
\caption{Fidelity $F_{r,n} $, Eq. \eqref{eq:fidelity}, as a function of the two-body binding energy $\E_r^{(2)}$. We focus here on the excited-state thresholds $r=1$ (top row), $r=2$ (center-top row) and $r=3$ (center-bottom row). Moreover, we display the overlap $|\langle\psi_{3,n}|\psi_{1,n}\rangle|^2$ (bottom row) between the three-body wave functions associated with the odd-numbered thresholds $r=1$ and $r=3$. The left and right columns distinguish between the case of two heavy bosons and two heavy fermions. The fidelities are presented for finite-range potentials of shape $f_\mathrm{L}$, Eq. \eqref{eq:fL} (filled red symbols) and $f_\mathrm{G}$, Eq. \eqref{eq:fG} (empty blue symbols). For a mass ratio of $M/m=20$ there are six three-body bound states associated to the individual heavy-light thresholds, three each for the case of bosons or fermions.% In all subfigures (a)-(h) the results for the two finite-range potentials are very close to each other. For $r=2$ the fidelities are close to unity and strictly monotonically increasing towards this limit value for $\E_r^{(2)}\to 0^-$. Here, the fidelities for fermions (d) are much higher and converging faster to unity in comparison to the ones obtained for bosons (c). In contrast, for the odd-numbered thresholds the fidelities are orders of magnitude below unity ((a), (b), (e), (f)). The overlap $|\langle\psi_{3,n}|\psi_{1,n}\rangle|^2$ is approaching unity, however no significant difference between bosons and fermions is visible ((g) and (h)).
}
\label{fig:fidelity}
\end{figure*}

In Fig. \ref{fig:fidelity} we present the fidelity $F_{r,n}$ as a function of the two-body binding energy $\E_r^{(2)}$. Throughout the subfigures we indicate by filled red and empty blue symbols the fidelities for a heavy-light interaction of shape $f_\L$, Eq. \eqref{eq:fL}, and $f_\mathrm{G}$, Eq. \eqref{eq:fG}, respectively. Analogous to Fig. \ref{fig:spectrum_R23}, the top three rows show the results for three different heavy-light thresholds $\E_1^{(2)}\to 0^-$, $\E_2^{(2)}\to 0^-$ and $\E_3^{(2)}\to 0^-$, respectively. In the last row we separately display the overlap $|\langle\psi_{3,n}|\psi_{1,n}\rangle|^2$. The left and right columns distinguish the case of two heavy bosons ($n=0,2,4$) or fermions ($n=1,3,5$), respectively.

A comparison of the fidelities between the two different finite-range heavy-light interactions shows a similar behavior as for the three-body energies. In all subfigures the corresponding red and blue data sets are very close to each other and thus suggest universality of the wave functions of three-body bound states.

Next, we analyze the universal behavior of the fidelities for different weakly-bound states in the heavy-light subsystems. In contrast to the results for the energies, which show the same universal limit for all values of $r$, we observe that the fidelities near the three excited-state thresholds display a fundamentally different behavior. A fidelity of unity indicates that the two wave functions overlap perfectly and are therefore identical. On the other hand a vanishing fidelity indicates that the two wave functions $\psi_{r,n}$ and $\psi_n^\star$ are orthogonal. For the even-numbered threshold ($r=2$) the fidelities $F_{2,n}$ are increasing and approaching unity as the threshold is approached, as demonstrated in Fig.~\ref{fig:fidelity} (c) and (d). Contrarily, for the odd-numbered thresholds ($r=1$ and $r=3$) the fidelities $F_{1,n}$ and $F_{3,n}$ are far below unity, as evident from Fig.~\ref{fig:fidelity} (a), (b), (e), and (f). However, the overlap $|\langle\psi_{3,n}|\psi_{1,n}\rangle|^2$ between the three-body wave functions near the odd-numbered thresholds approaches again unity as $\E_r^{(2)} \to 0^-$, see Fig. \ref{fig:fidelity} (g) and (h).

To conclude, these results for the fidelities indicate that the three-body wave functions approach two distinct universal limits, one for symmetric, and another one for antisymmetric weakly-bound heavy-light states. The universal limits for the symmetric case are those for the zero-range contact interaction. On the other hand, the universal limits for the antisymmetric case might correspond to those obtained for an odd-wave pseudopotential~\cite{Girardeau2004}. We recall that such a dependence on the symmetry of the weakly-bound heavy-light state has not become apparent in the universal limit of the respective three-body energies.

Finally, we analyze the universal behavior for different excited three-body states characterized by the index $n$. Near the even-numbered threshold ($r=2$), the fidelities $F_{2,n}$ are larger for fermionic heavy particles (odd $n$) than for bosonic ones (even $n$) and the limit value for $\E_2^{(2)}\to 0^-$ is reached faster. This is in accordance with the results for the energies shown in Fig. \ref{fig:spectrum_R23}. In contrast, near the odd-numbered thresholds ($r=1$ and $r=3$) the overlap $|\langle\psi_{3,n}|\psi_{1,n}\rangle|^2$ shows no significant difference between bosons and fermions. The fidelities $F_{1,n}$ and $F_{3,n}$ are far below unity and therefore we refrain from analyzing them in more detail.

\begin{figure*}[htbp]
   \centering
   \subfigure[\hspace{\columnwidth}]{
\begin{tabular}{lcc}
&\multicolumn{2}{c}{$n=0$ (Bosons)} \\
&$\abs{\E_r^{(2)}}=10^{-3}$ &$\abs{\E_r^{(2)}}=10^{-7}$ \\
\rotatebox[origin=c]{90}{$r=0$}&
$\vcenter{\hbox{\includegraphics[width=.21\textwidth]{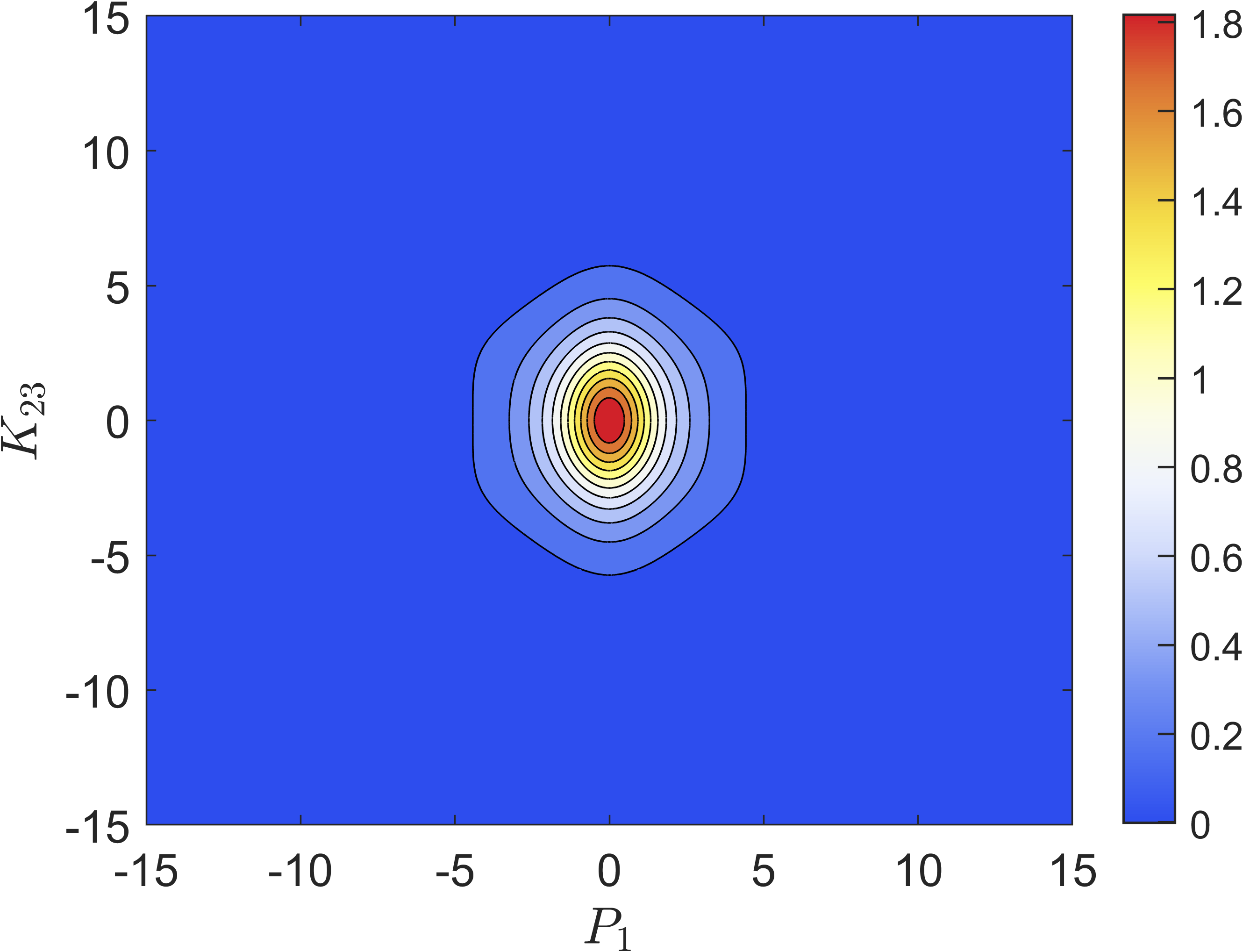}}}$&
$\vcenter{\hbox{\includegraphics[width=.21\textwidth]{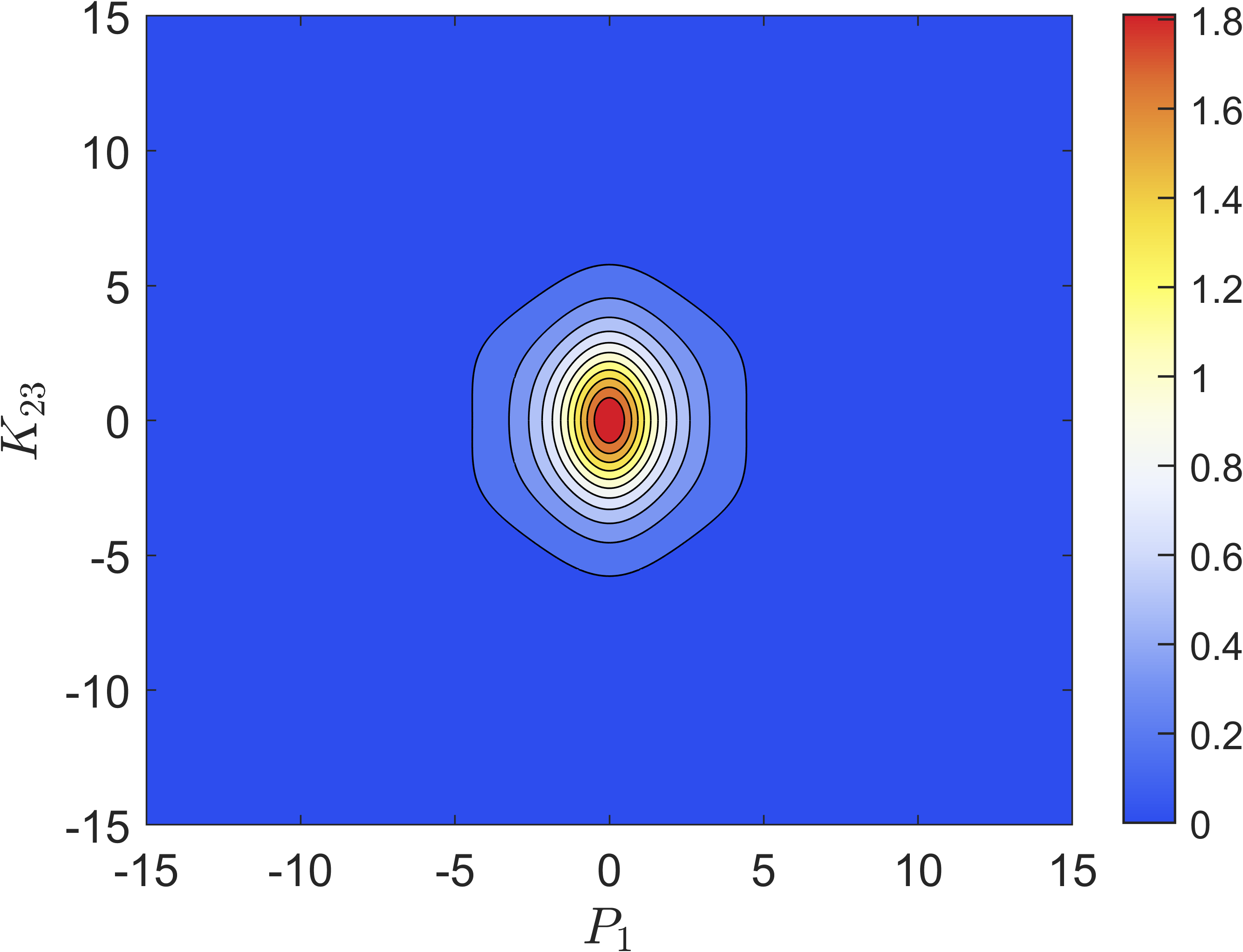}}}$\\
\\
\rotatebox[origin=c]{90}{$r=1$}&
$\vcenter{\hbox{\includegraphics[width=.21\textwidth]{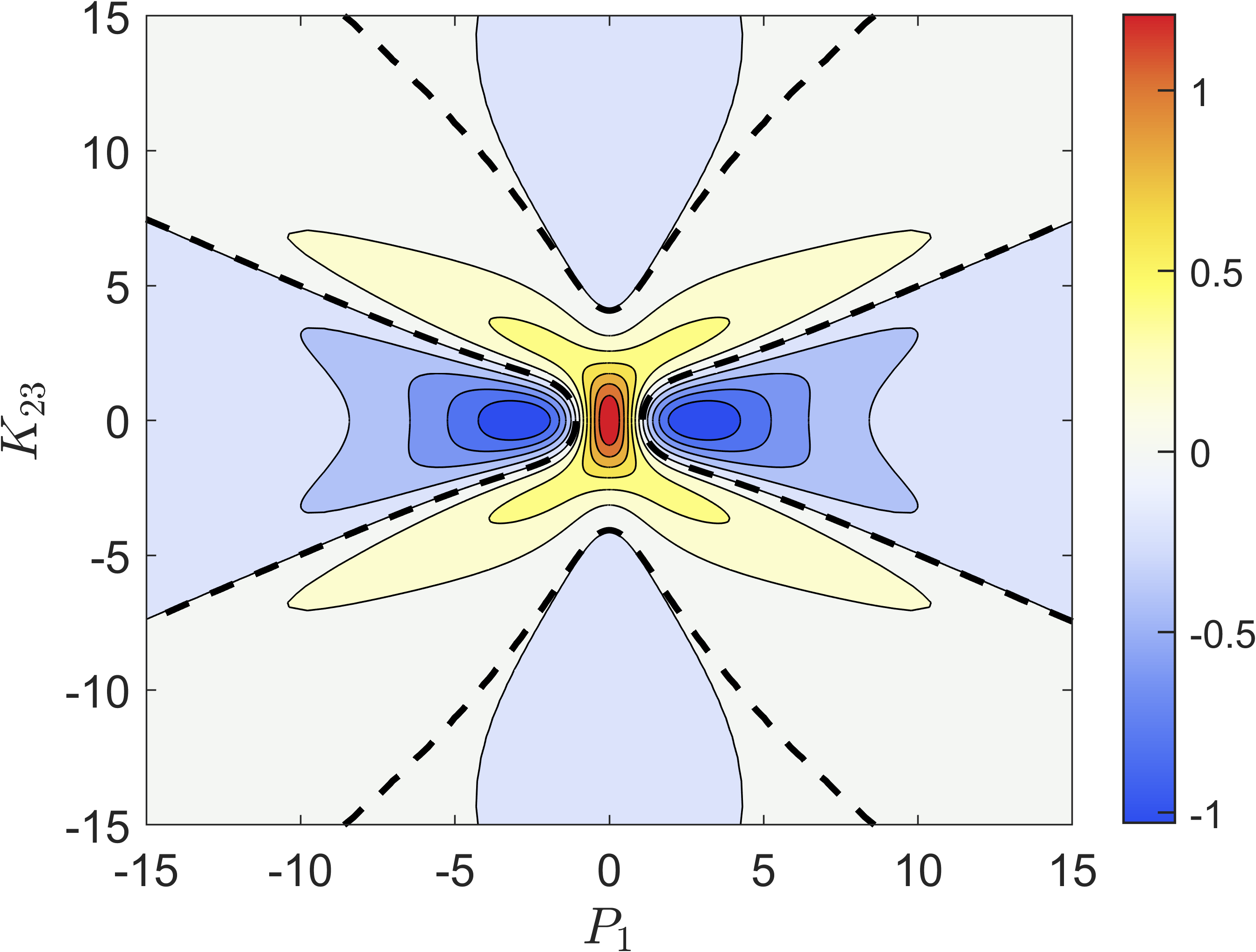}}}$&
$\vcenter{\hbox{\includegraphics[width=.21\textwidth]{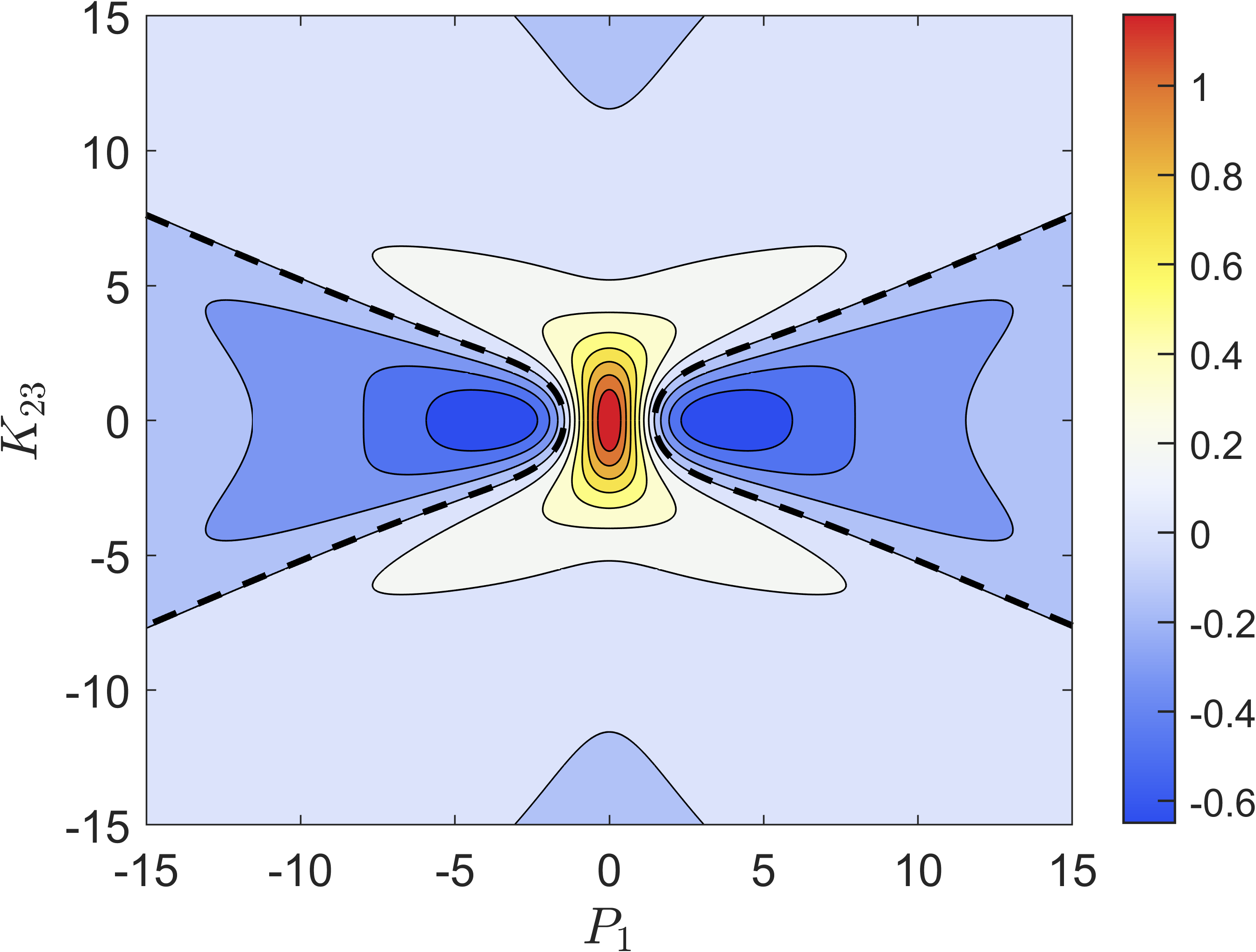}}}$\\
\\
\rotatebox[origin=c]{90}{$r=2$}&
$\vcenter{\hbox{\includegraphics[width=.21\textwidth]{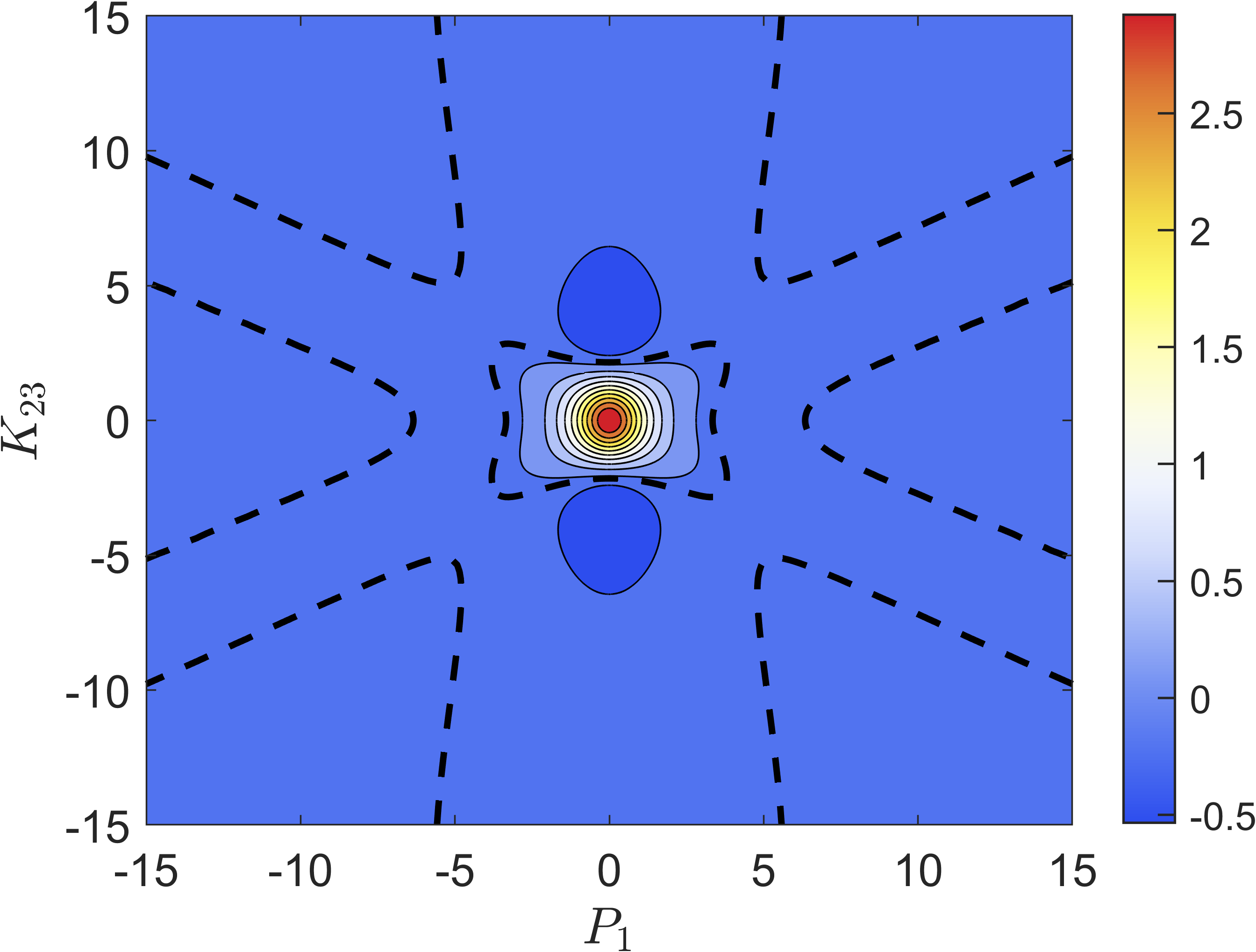}}}$&
$\vcenter{\hbox{\includegraphics[width=.21\textwidth]{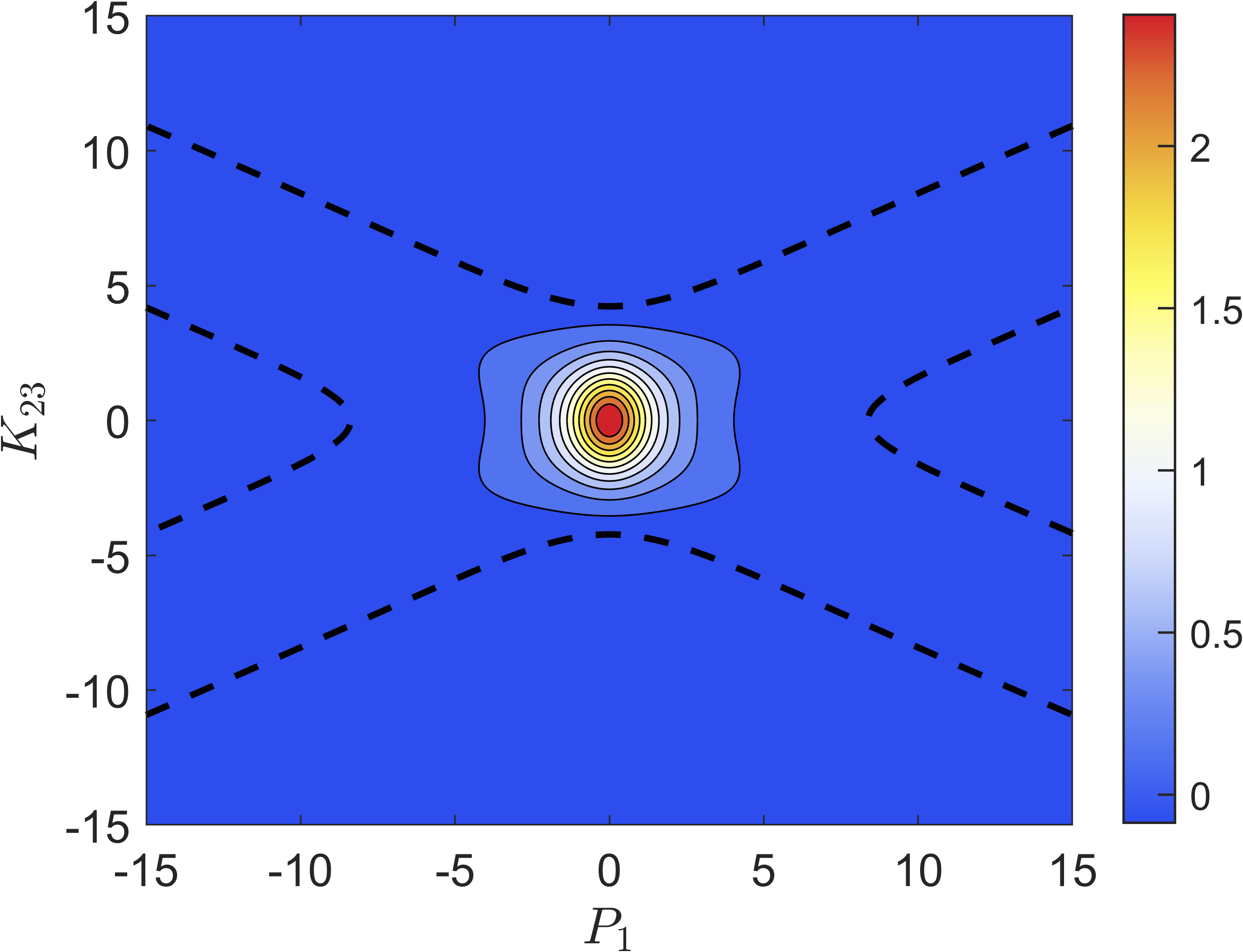}}}$\\
\\
\rotatebox[origin=c]{90}{$r=3$}&
$\vcenter{\hbox{\includegraphics[width=.21\textwidth]{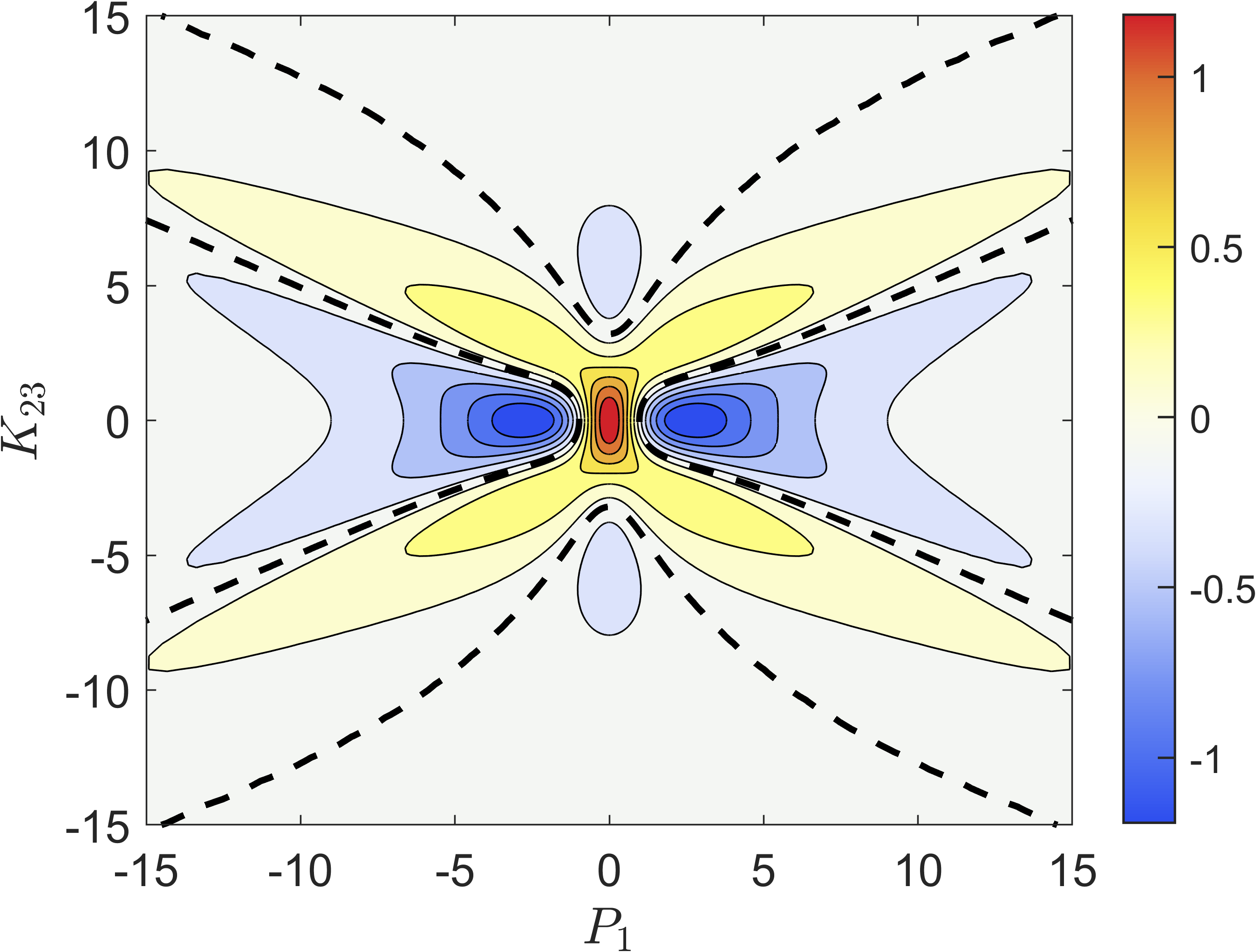}}}$&
$\vcenter{\hbox{\includegraphics[width=.21\textwidth]{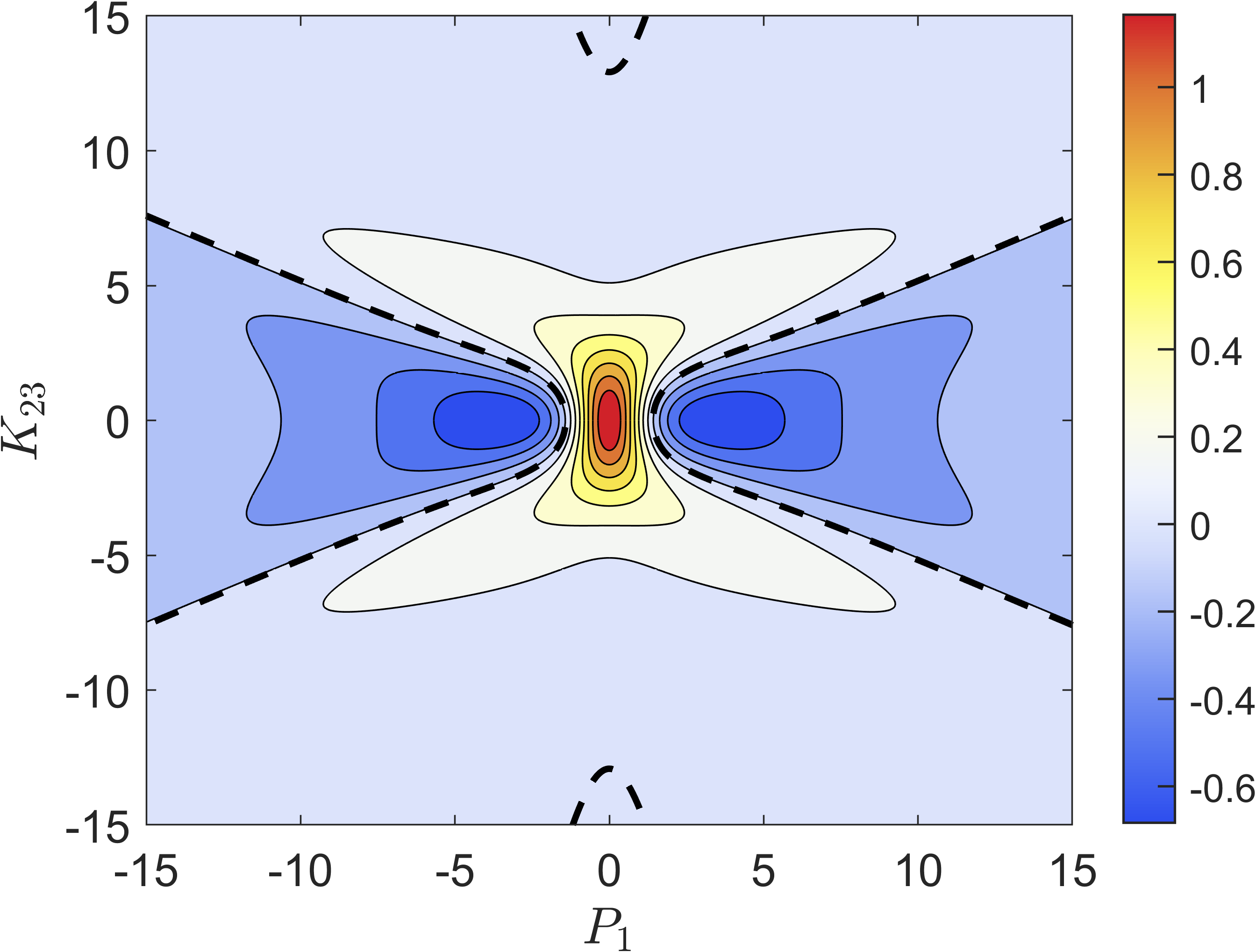}}}$
\end{tabular}}
\hfill
   \subfigure[\hspace{\columnwidth}]{
\begin{tabular}{lcc}
&\multicolumn{2}{c}{$n=1$ (Fermions)} \\
&$\abs{\E_r^{(2)}}=10^{-3}$ &$\abs{\E_r^{(2)}}=10^{-7}$ \\
\rotatebox[origin=c]{90}{$r=0$}&
$\vcenter{\hbox{\includegraphics[width=.21\textwidth]{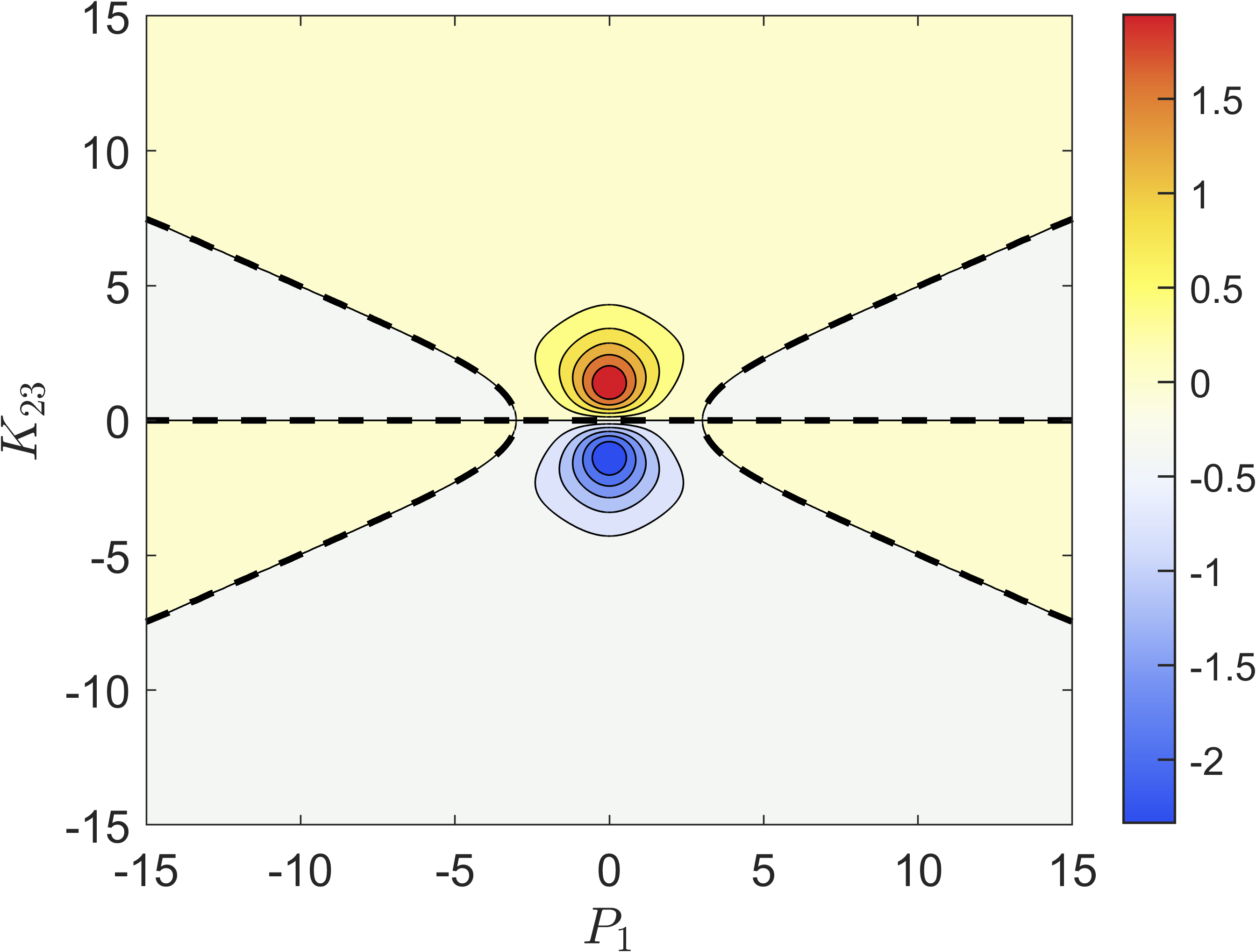}}}$&
$\vcenter{\hbox{\includegraphics[width=.21\textwidth]{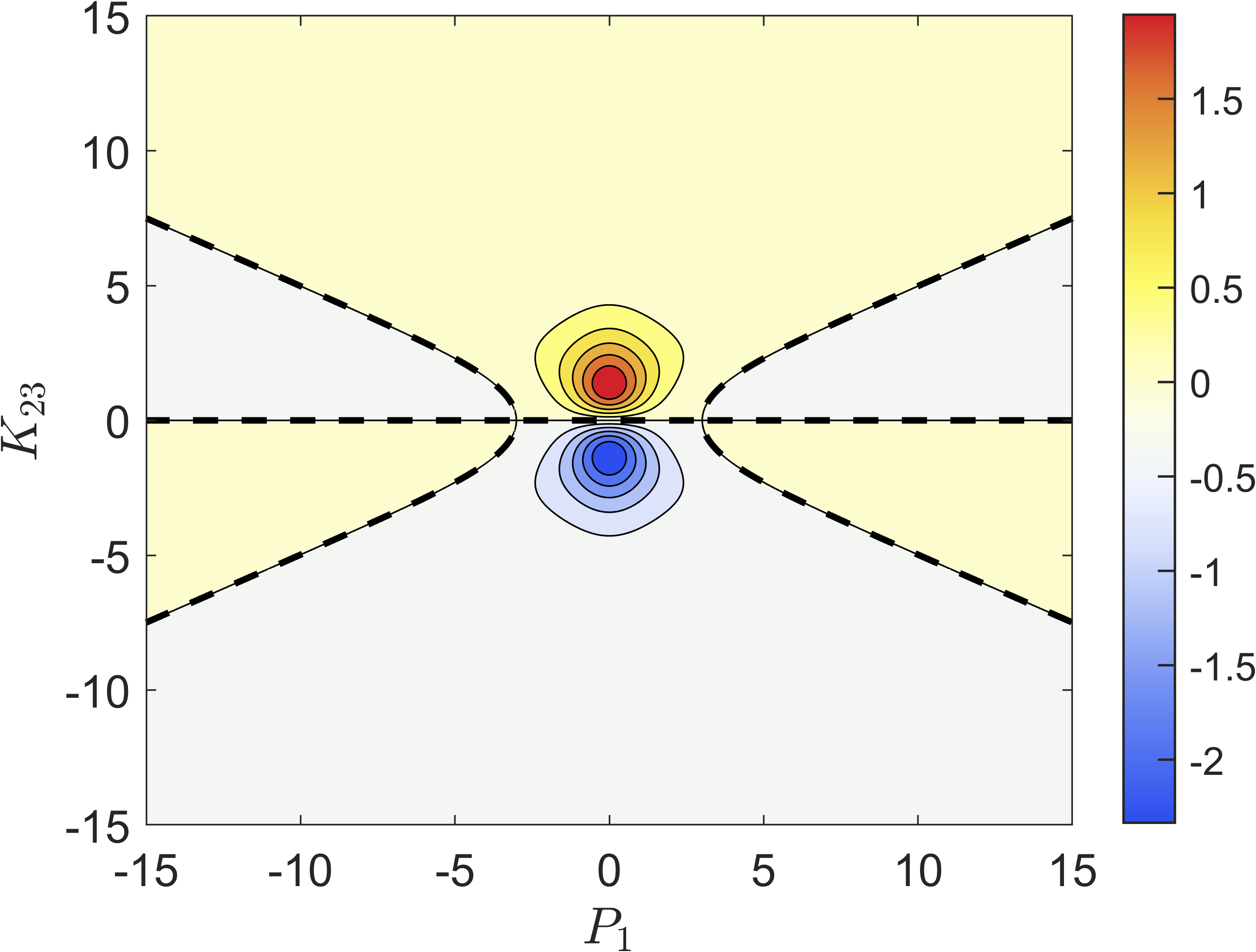}}}$\\
\\
\rotatebox[origin=c]{90}{$r=1$}&
$\vcenter{\hbox{\includegraphics[width=.21\textwidth]{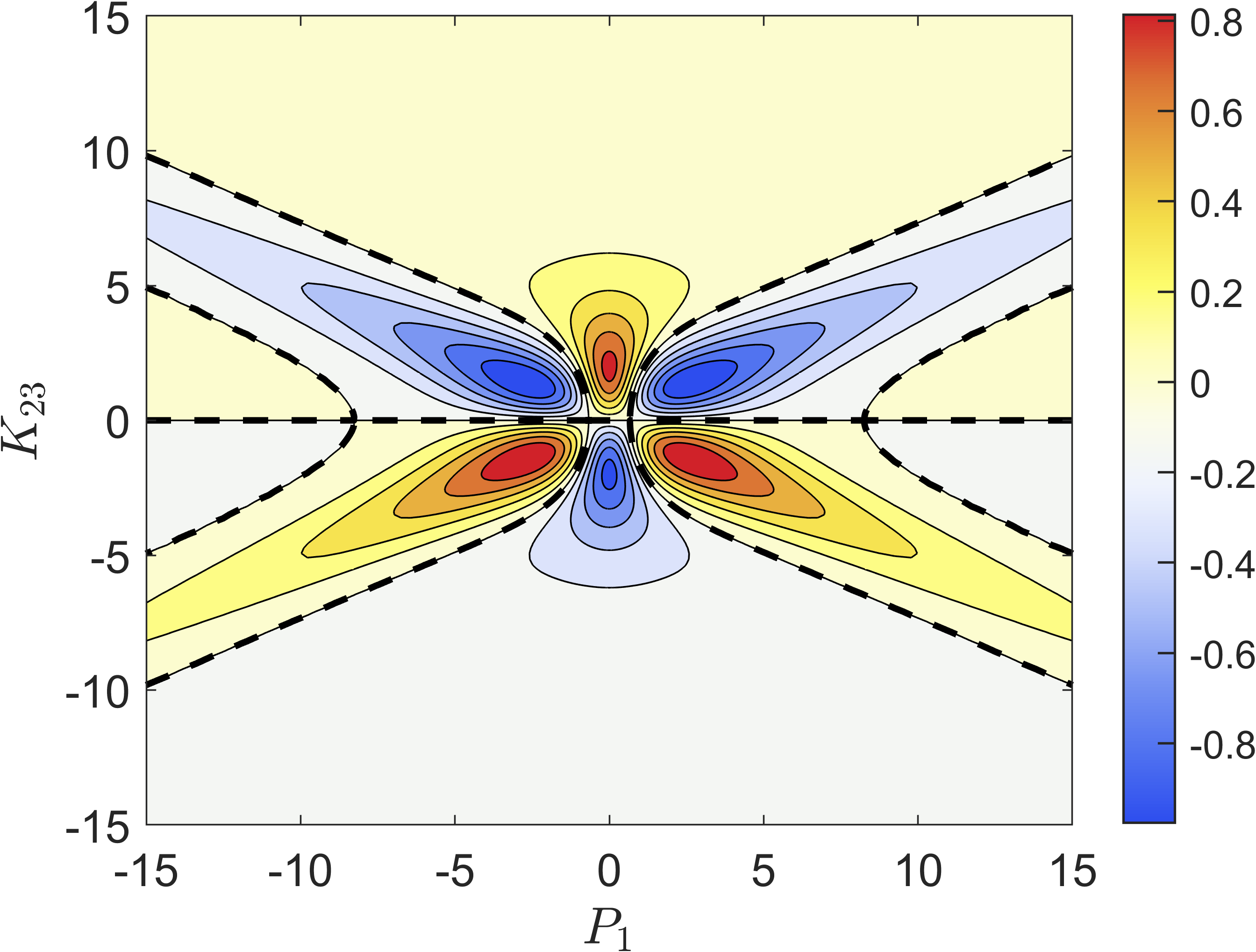}}}$&
$\vcenter{\hbox{\includegraphics[width=.21\textwidth]{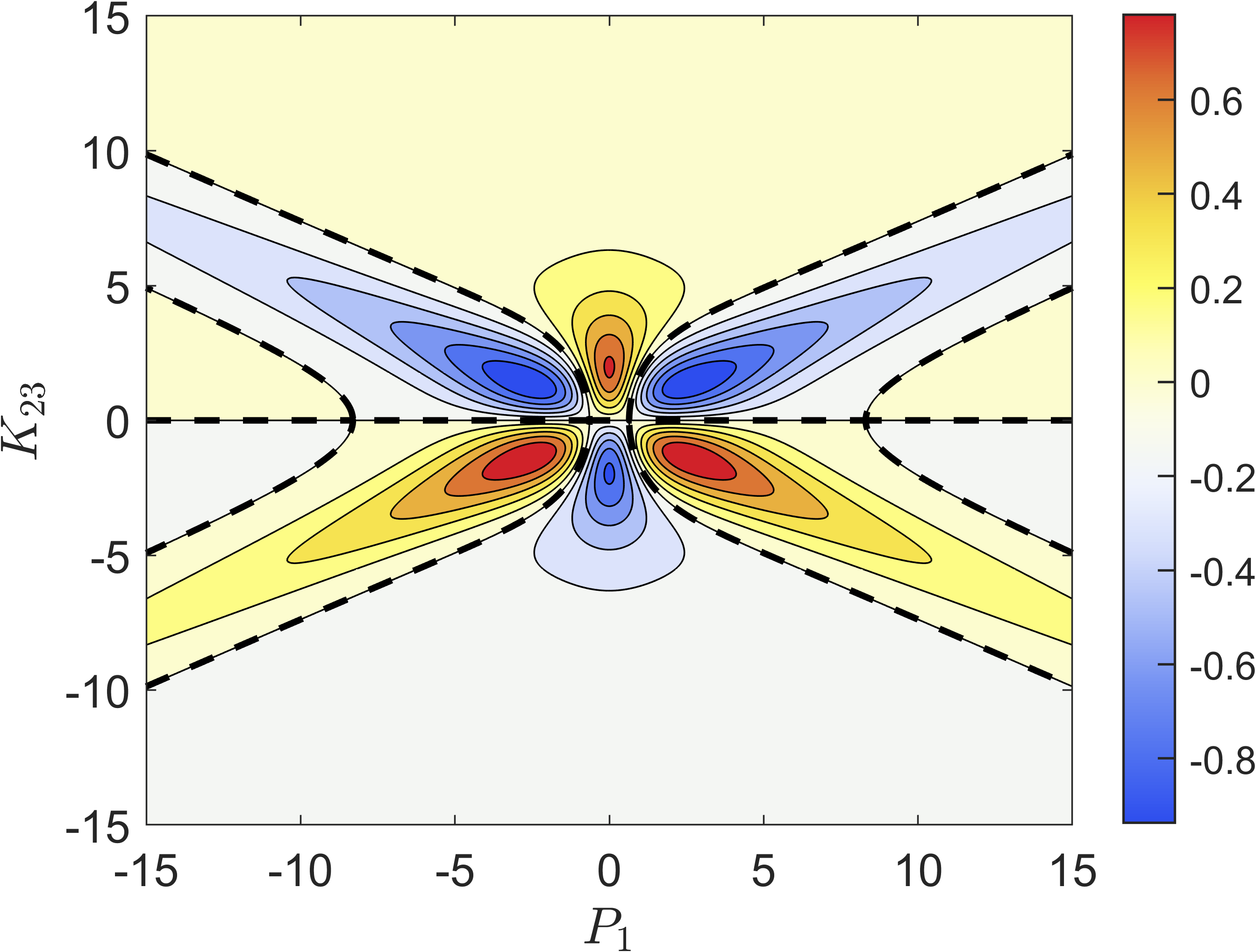}}}$\\
\\
\rotatebox[origin=c]{90}{$r=2$}&
$\vcenter{\hbox{\includegraphics[width=.21\textwidth]{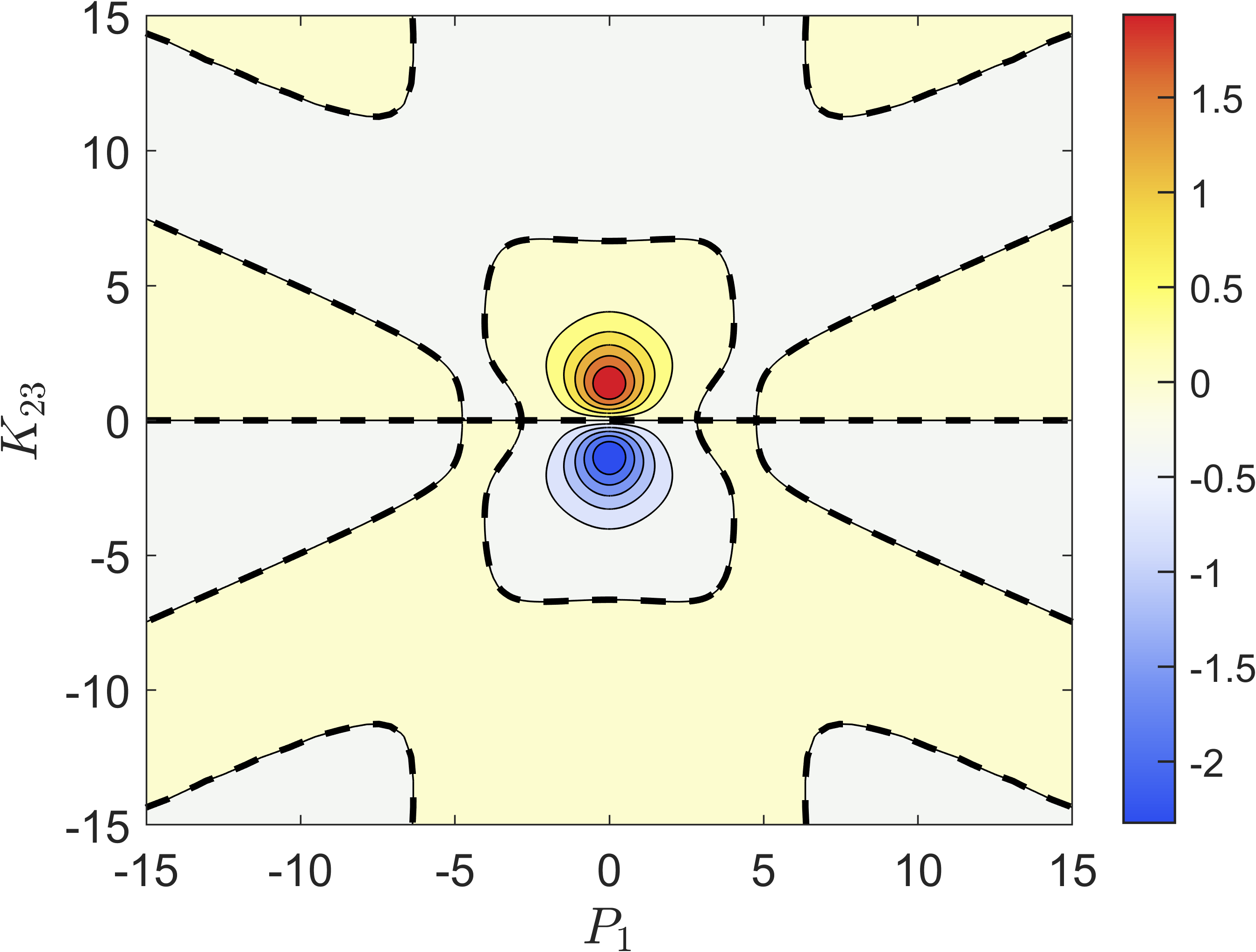}}}$&
$\vcenter{\hbox{\includegraphics[width=.21\textwidth]{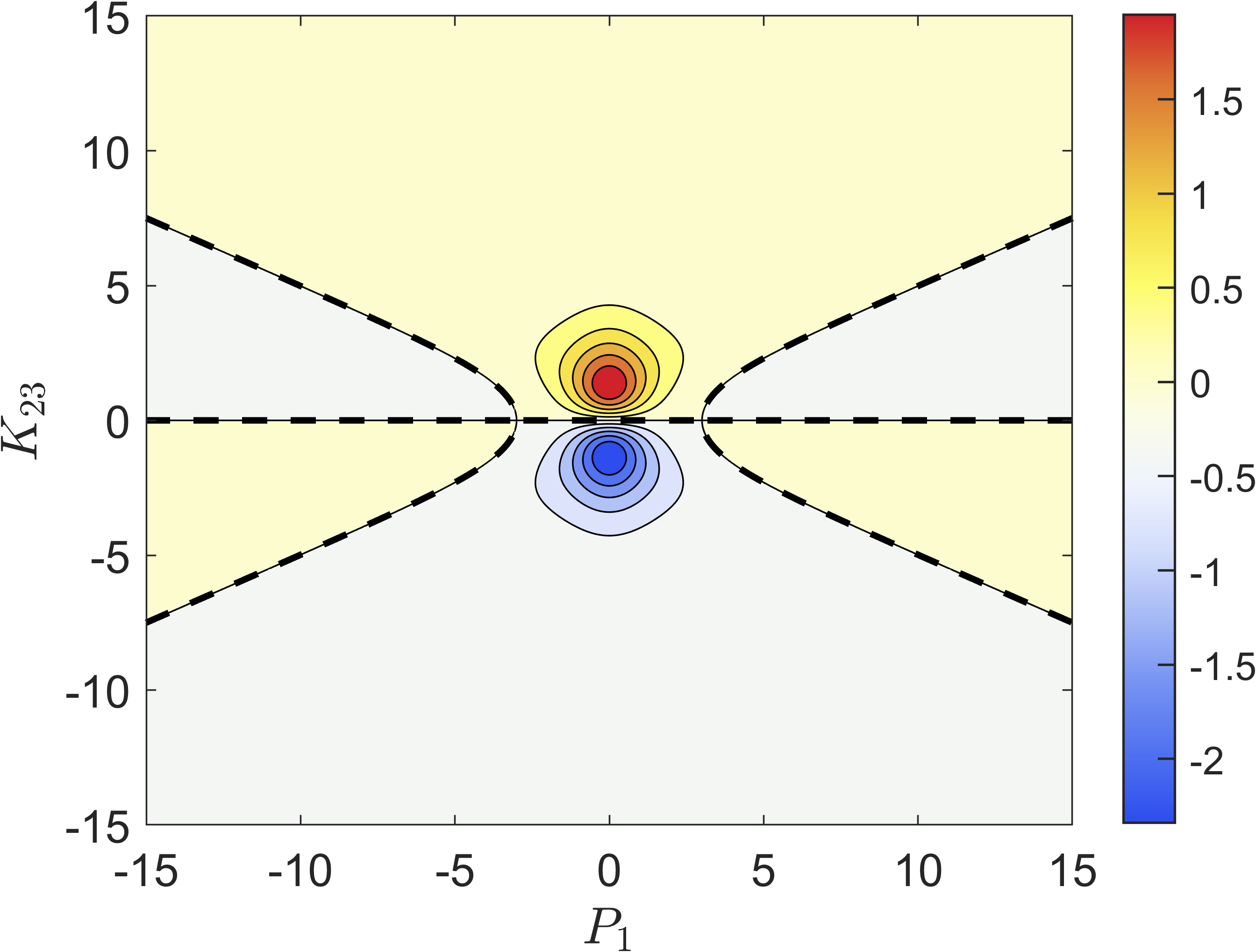}}}$\\
\\
\rotatebox[origin=c]{90}{$r=3$}&
$\vcenter{\hbox{\includegraphics[width=.21\textwidth]{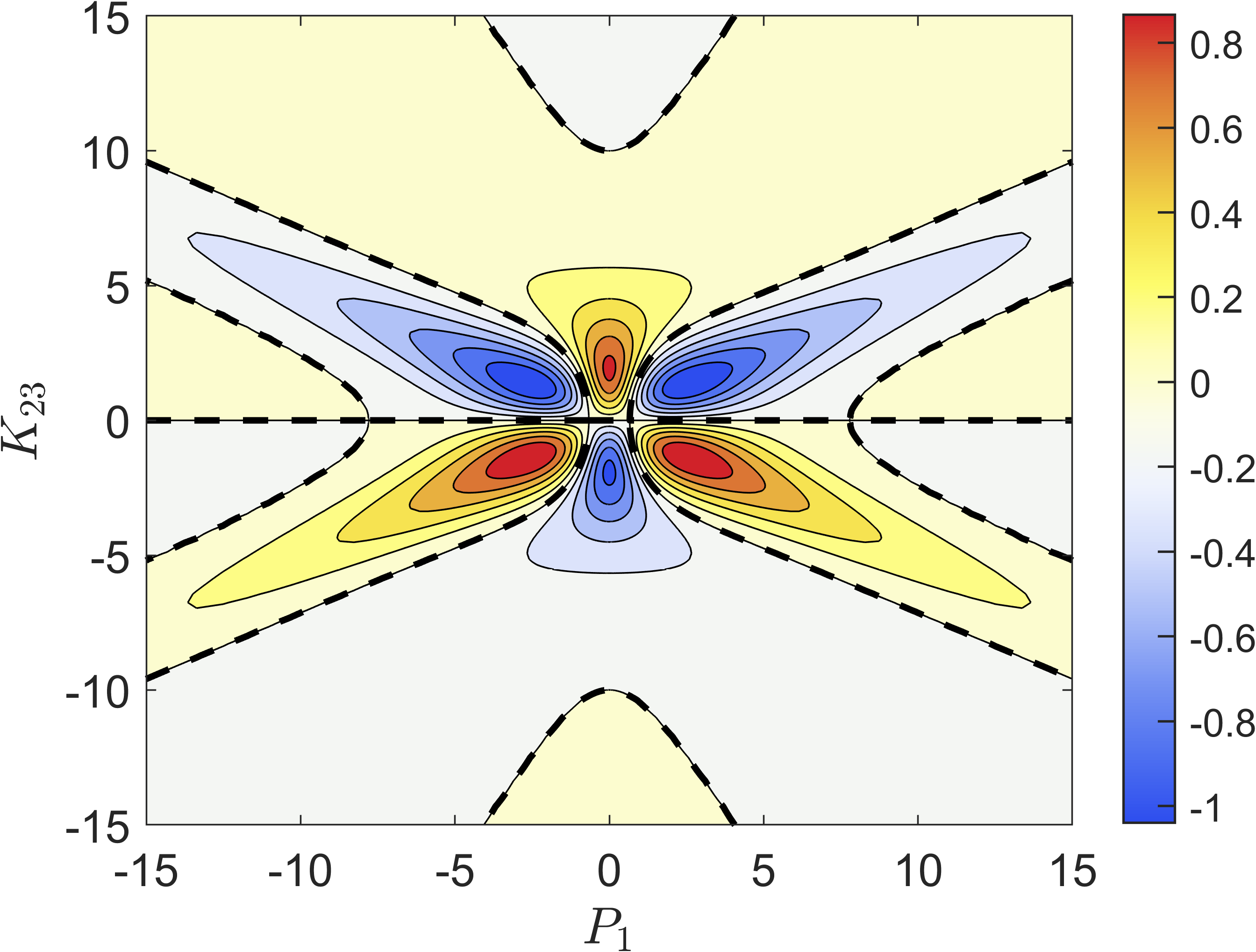}}}$&
$\vcenter{\hbox{\includegraphics[width=.21\textwidth]{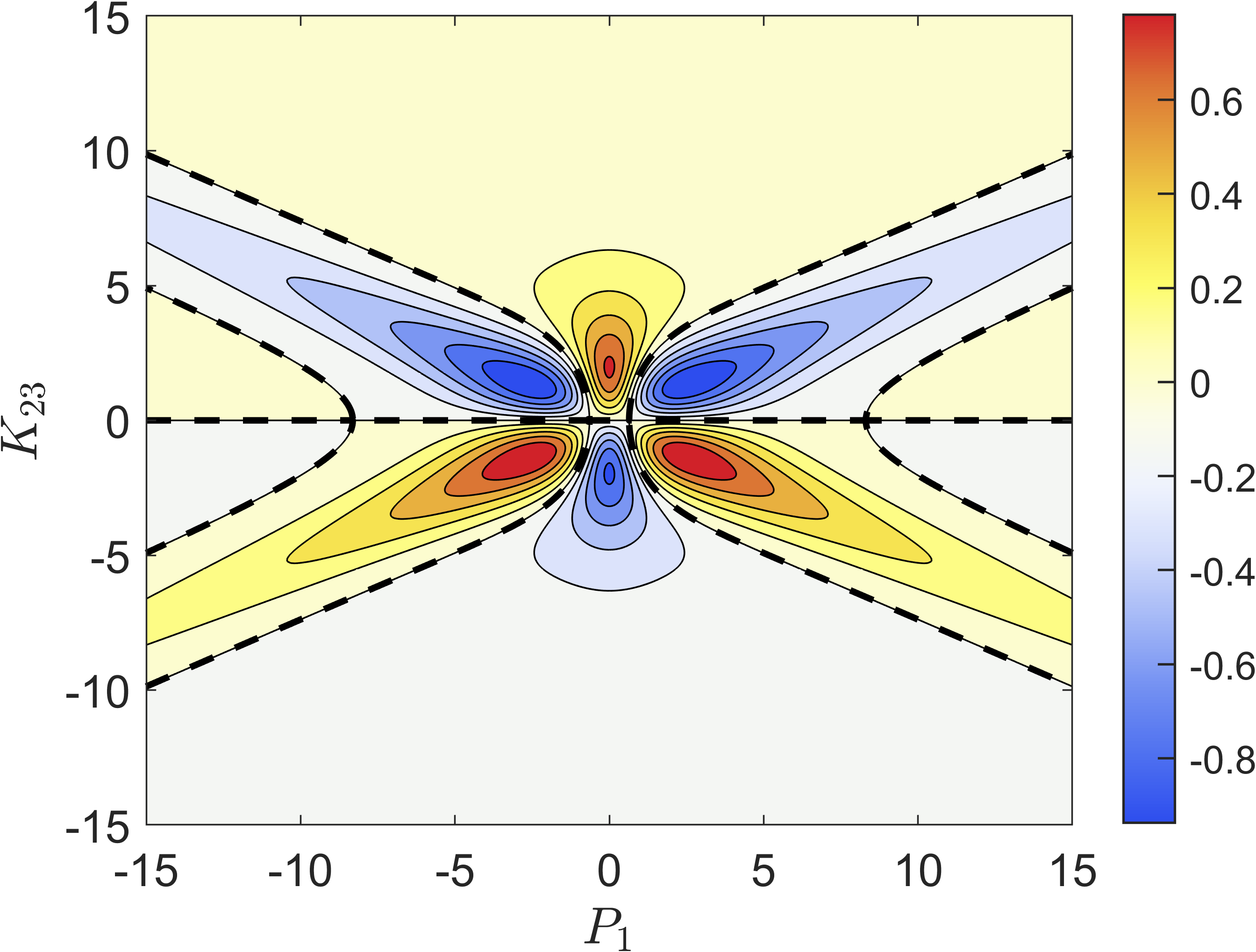}}}$
\end{tabular}}
\caption{Contour plots of the normalized three-body wave functions $\psi_{r,n}(K_{23},P_1)$ in the space spanned by the scaled momenta, Eq. \eqref{eq:scaledmomenta}, as obtained for the heavy-light interaction of shape $f_\G$, Eq. \eqref{eq:fG}. The rows distinguish between the different two-body thresholds, $r=0$ (top), $r=1$ (center-top), $r=2$ (center-bottom) and $r=3$ (bottom). The diagrams are split into two halves, corresponding to (a) bosons ($n=0$) and (b) fermions ($n=1$). Each half is split again into two columns, the left column in each half shows the case of $\abs{\E_r^{(2)}}=10^{-3}$, and the right column the case of $\abs{\E_r^{(2)}} = 10^{-7}$. The contours where the wave functions vanish are highlighted by dashed black lines.}
\label{fig:ContourPlots}
\end{figure*}

In order to reinforce our results, we display contour plots of the wave functions in Fig. \ref{fig:ContourPlots}. We restrict ourselves here to the potential shape $f_\G$, Eq. \eqref{eq:fG}, of the heavy-light interaction and to the most deeply bound states for the case of bosons ($n=0$) and fermions ($n=1$). The wave functions are displayed in momentum space
\begin{equation}\label{eq:scaledmomenta}
P_{1} \equiv \frac{p_{1}}{\sqrt{2\abs{\E_r^{(2)}}}}\, , \qquad K_{23} \equiv \frac{k_{23}}{\sqrt{2\abs{\E_r^{(2)}}}}
\end{equation}
scaled by the two-body energy $\E_r^{(2)}$ of the corresponding weakly-bound heavy-light state.

The four rows distinguish the two-body thresholds $r=0$ (top), $r=1$ (center-top), $r=2$ (center-bottom) and $r=3$ (bottom). The diagrams are split into two halves, the left one (a) corresponds to bosons ($n=0$) whereas the right one (b) corresponds to fermions ($n=1$). Each half is split again into two columns, the left column in each half displays the wave functions for the two-body binding energy $\abs{\E_r^{(2)}}=10^{-3}$, and the right column for $\abs{\E_r^{(2)}} = 10^{-7}$. The contours of zero value for the wave functions are highlighted by black dashed lines.

First we discuss the even-numbered thresholds. The three-body wave functions near the ground state threshold $r=0$ (top) and the excited-state threshold $r=2$ (center-bottom) look already very similar for the respective state $n$. This similarity is further enhanced when the two-body threshold is approached. This behavior is in line with the fidelities increasing towards unity as presented in Fig. \ref{fig:fidelity} (c) and (d). Hence, from this we deduce that the three-body wave functions for the even-numbered heavy-light thresholds approach the same universal limit.

Next, we consider the odd-numbered thresholds. As presented in Fig. \ref{fig:ContourPlots}, the three-body wave functions for $r=1$ (center-top) and $r=3$ (bottom) are very similar, but look fundamentally different from those obtained near the even-numbered thresholds $r=0$ (top) and $r=2$ (center-bottom). This is in agreement with the high overlap shown in Fig. \ref{fig:fidelity} (g) and (h) and the low values of the corresponding fidelities (Fig. \ref{fig:fidelity} (a), (b), (e), (f)). As a result, this expresses that the same universal limit of the three-body wave functions is approached near all odd-numbered heavy-light thresholds. However, this limit is distinct from the limit near even-numbered thresholds.

Subsequently, we analyze the three-body wave functions depending on the index $n$ in order to highlight differences and similarities between the cases of two heavy bosons ($n=0$) and fermions ($n=1$). This property of the two heavy particles can be identified for all values of $r$ from the symmetry with respect to the line $K_{23}=0$, as evident from Fig.~\ref{fig:ContourPlots} and summarized by Eq.~\eqref{eq:bosferm_MainText}. As pointed out already in section \ref{sec:methods}, this statement translates in coordinate space to the same symmetry with respect to the exchange $y_{23}\to -y_{23}$. As an example, the three-body wave functions in coordinate representation are displayed in Ref. \cite{Happ2019} for the contact interaction ($r=0$). However, already from Fig. \ref{fig:ContourPlots} it is apparent that the bosonic wave functions have parity $+1$, whereas the fermionic ones have parity $-1$, which is summarized by Eq. \eqref{eq:bosferm_MainText}.

For these two different kinds of heavy particles, we now discuss the speed of convergence towards a scale-invariant three-body bound state when a particular two-body threshold is approached. With scale-invariance we mean here that the wave functions expressed in the scaled momenta, Eq. \eqref{eq:scaledmomenta}, behave like a constant as a function of $\E_r^{(2)}$. Near the ground-state threshold ($r=0$) both bosonic (Fig.~\ref{fig:ContourPlots} (a)) and fermionic (Fig.~\ref{fig:ContourPlots} (b)) wave functions look identical for the presented two-body binding energies $|\mathcal{E}_r^{(2)}|=10^{-3}$ and $|\mathcal{E}_r^{(2)}|=10^{-7}$. This is because they have already reached the scale-invariant limit \cite{Happ2019}. However, for the excited-state thresholds ($r>0$), the structure of the bosonic wave functions (Fig.~\ref{fig:ContourPlots} (a)) still changes quite significantly when decreasing the two-body binding energy. On the other hand, the fermionic wave functions (Fig.~\ref{fig:ContourPlots} (b)) are already much closer to the scale-invariant regime. This behavior of the wave functions also enables a better understanding of the results presented in Fig. 3. Indeed, the fidelities for $r=2$ are lower for bosons, Fig.~\ref{fig:fidelity} (c), than for fermions, Fig.~\ref{fig:fidelity} (d). Consequently, we observe a major difference between bosons and fermions in their speed of convergence towards the scale-invariant regime. This is in line with the results for the energy spectrum, discussed in subsection \ref{sec:EnergySpectrum}. In conclusion, we note that the difference in the wave functions having an extremum (bosons, Fig.~\ref{fig:ContourPlots} (a)) or zero (fermions, Fig.~\ref{fig:ContourPlots} (b)) at the line $K_{23}=0$ might explain this faster convergence of the energies and wave functions for the fermionic case in comparison to the bosonic one.

Finally, we address another aspect of the universal behavior of the three-body bound states. Indeed, in each column of Fig.~\ref{fig:ContourPlots}, that is for fixed two-body binding energies $\E_{r}^{(2)}$, the wave functions for both bosons and fermions associated with the odd-numbered thresholds ($r=1$ and $r=3$) look very similar. We point out that this is true although the bosonic ones have not yet reached the scale-invariant regime. This also explains why the overlap $|\langle\psi_{3,n}|\psi_{1,n}\rangle|^2$ between the three-body wave functions near odd-numbered thresholds does not display a significant difference as shown in Fig.~\ref{fig:fidelity} (g) for the bosonic, and in Fig.~\ref{fig:fidelity} (h) for the fermionic case. We expect the same behavior when comparing the results for any two excited-state thresholds for which the associated weakly-bound states have the same symmetry, \textit{i.e.} $r=2$ and $r=4$.

\section{Influence of deeply bound two-body states on the three-body universality}\label{sec:deeplybound}
In the preceding section we have found universality in the three-body system, when the heavy-light subsystems are tuned towards an excited-state threshold. In particular, the universality of the three-body wave functions depends on the symmetry of the weakly-bound state in the heavy-light subsystems. A natural question therefore arises whether the universal behavior is solely determined by this weakly-bound state. 

In order to answer this question we analyze in this section the influence of deeply bound heavy-light states on the universal limit in the three-body system. These deeply bound states have a nonzero binding energy and are naturally implied when the heavy-light subsystems are tuned close to an excited-state threshold. We focus our attention here on the energy ratios $\epsilon_{r,n}$, Eq. \eqref{epsilon}.

The separable expansion \cite{Weinberg1963,Sitenko1991}, outlined in Appendix~\ref{app:separableexpansion}, is well suited for this analysis. Indeed, near a particular threshold $\E_r^{(2)}\to 0^-$, each term in Eq.~\eqref{eq:sep_phi_MainText} with $\nu\leq r$ is related to a deeply bound state in the heavy-light subsystem, provided the energy arguments $\E_q$ coincide with the energy $\E_\nu^{(2)}$ of the respective two-body state. Similarly, for $\E_q = \E_r^{(2)}$, the term $\nu =r$ describes the weakly-bound two-body bound state. Thus, the influence of the deeply bound two-body states on the universality of three-body bound states associated to a particular threshold ($r>0$) can be tested by manually including and excluding different terms with $\nu<r$.

\begin{figure}[htb]
\centering
\includegraphics[width=\columnwidth]{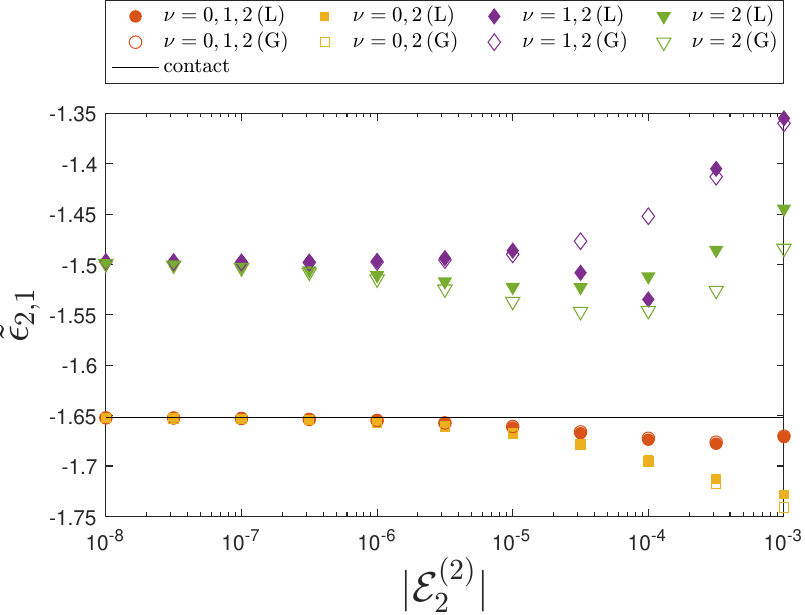}
\caption{Ratio $\tilde{\epsilon}_{2,1}$ of three-body binding energy to $\E_2^{(2)}$ for $r=2$, $n=1$ and different numbers of separable terms in Eqs.~\eqref{eq:sep_phi_MainText} and \eqref{eq:FE_final_MainText} as a function of $\E_2^{(2)}$. We manually include or exclude different expansion terms as indicated in the legend. Filled symbols display the results for the two-body interaction of shape $f_\L$, Eq. \eqref{eq:fL}, and empty symbols for the shape $f_\G$, Eq. \eqref{eq:fG}, accordingly. The exact result for the energy ratio $\epsilon_{2,1}$ in the limit $\E_2^{(2)} \to 0^-$ is indicated by a black line.}
\label{fig:deepstatesareimportant}
\end{figure}

Here we consider as an example the case $r=2$ where there are two ($\nu = 0$ and $\nu = 1$) deeply bound states in the heavy-light system in addition to the weakly-bound state with $\nu=2$. Moreover, the fermionic case ($n=1$) is chosen because of its faster convergence in the limit of vanishing two-body binding energy $\E_r^{(2)}$ compared to the bosonic case. In Fig. \ref{fig:deepstatesareimportant} we illustrate the influence of the deeply bound states on the energy ratio $\tilde{\epsilon}_{r,n}$.  This ratio converges towards $\epsilon_{r,n}$, Eq. \eqref{epsilon}, provided that all expansion terms are included in Eqs. \eqref{eq:sep_phi_MainText} and \eqref{eq:FE_final_MainText}. To explore the influence of particular expansion terms on the three-body energy, we consider different combinations in the expansion as indicated by the legend. Filled symbols mark the results for an interaction potential of shape $f_\L$, Eq. \eqref{eq:fL}, and empty ones for the shape $f_\G$, Eq. \eqref{eq:fG}. We consider four different combinations:

(i) ($\nu = 0,1,2$; red dots) If both deeply bound two-body states and the weakly-bound one are included, the ratio $\tilde{\epsilon}_{2,1}$ converges to the limit value $\epsilon_{1}^\star$ for the contact interaction (black line).

(ii) ($\nu = 0,2$; yellow squares) Excluding the first excited heavy-light bound state ($\nu = 1$) from the analysis has no observable impact on the limit value of the energy ratio.

(iii) ($\nu = 1,2$; purple diamonds) When excluding instead the heavy-light ground state, described by the term $\nu = 0$, the energy ratio converges to a limit value that is different from $\epsilon_1^\star$ of the contact interaction.

(iv) ($\nu = 2$; green triangles) The exclusion of both deeply bound states with $\nu = 0$ and $\nu =1$ leads to the same (incorrect) limit value as for (iii).

As evident from Fig. \ref{fig:deepstatesareimportant} we observe that for  all combinations (i)-(iv) the limit values for the two finite-rage potentials of shape $f_\L$ and $f_\G$ coincide. This interaction-independent behavior is not visibly influenced by the presence or absence of the deeply bound two-body states. Hence, it is the origin for the universality of the energy ratios $\epsilon_{r,n}$, Eq. \eqref{epsilon}, presented in section \ref{sec:Results}.

On the other hand, the separable approximation, that is taking into account only the term $\nu=r=2$ corresponding to the weakly-bound state of the heavy-light subsystem, yields a value of $\tilde{\epsilon}_{r,n}$ that does not coincide with $\epsilon_{2,1}$. Therefore, the weakly-bound state alone is \textit{not} sufficient to achieve an agreement with the universal limit for the ground-state two-body threshold. Hence, the result presented in section \ref{sec:Results}, that the three-body energies approach the same universal limit for all values of $r$, crucially relies on the presence of the deeply bound two-body states and cannot be explained by the weakly-bound state $\nu =r =2$ alone. However, it seems that not all deeply bound states are equally relevant. For the example presented here, it is the term $\nu = 0$ that is important, whereas the term $\nu =1$ has negligible effect on $\tilde{\epsilon}_{2,1}$ in the limit $\E_2^{(2)} \to 0^-$.

As a result, this analysis indicates that the separable approximation fails in predicting the correct limiting three-body energy ratios, whenever the heavy-light subsystems support deeply bound two-body states ($r>0$). However, in this article we have shown that the correct three-body energy ratios converge to the ones for the contact interaction. Indeed, the universal three-body energies for the finite-range potentials can be obtained when taking into account all deeply-bound states. A similar study on the separable approximation has been performed in Ref.~\cite{Mestrom2019} for three identical bosons in 3D.

\section{Conclusion and Outlook}\label{sec:conclusion}
In this article we have investigated the universality of a quantum mechanical heavy-heavy-light system  constrained to one dimension, provided the heavy-light subsystem is tuned to the energy threshold associated with a weakly-bound excited state. In comparison to the case of the ground-state threshold \cite{Happ2019}, the situation here differs in two main points: (i) the wave function of the weakly-bound heavy-light state has at least one node and its symmetry can change from symmetric to antisymmetric, and (ii) the heavy-light subsystems now contain additional deeply bound states. We have analyzed  the influence of both aspects on the three-body system by solving the Faddeev equations numerically for different finite-range potentials. Moreover, we have compared the resulting energies and wave functions to the corresponding results near the ground-state heavy-light threshold, as represented by those for the contact interaction.

Universality describes the property that a three-body quantity becomes independent of the short-range details of the two-body interaction potential. Since two finite-range potentials with different long-range decay have yielded the same three-body binding energies and wave functions, we have found an indication of universality for these two quantities. In particular, we have demonstrated this behavior near the thresholds corresponding to the first three excited bound states in the heavy-light subsystem. For the ground-state threshold, universality has already been demonstrated and proven in Ref. \cite{Happ2019}.

We have further compared the universality of the three-body system for the four heavy-light thresholds characterized by $r=0,1,2,3$. For the three-body binding energies, we have observed the same universal limit across all four thresholds. In particular, this suggests that the universal limits of the three-body binding energies do not depend on the symmetry of the weakly-bound heavy-light state. No fundamentally different result is expected near thresholds of higher excited ($r>3$) heavy-light states. However, we have found that for $r>0$ the speed of convergence towards the universal limit values is slower compared to $r=0$. In terms of the corresponding three-body wave functions, we have observed two distinct universal limits, one for all symmetric ($r=0,2,\ldots$), and a different one for all antisymmetric ($r=1,3,\ldots$) weakly-bound heavy-light states.

In addition, the universal behavior depends strongly on the property of the two heavy particles being bosons or fermions. In particular, close to an excited-state threshold the three-body energy spectrum and corresponding wave functions in the fermionic case display a scale-invariant behavior at much larger two-body binding energies than in the bosonic one. This is in contrast to the ground-state threshold, where no such difference between bosonic and fermionic heavy particles was observed \cite{Happ2019}. We want to highlight here that the two heavy particles being bosons or fermions does not affect the properties of the heavy-light subsystems. Any observed difference in the three-body quantities between the two cases is therefore a pure three-body effect.

Furthermore, we have demonstrated that for finite-range heavy-light interactions tuned close to an excited-state threshold, the separable approximation of the corresponding interactions does not provide the correct universal limits of the three-body energies. This has revealed that in this case the universal limit of the three-body binding energies cannot be explained with the weakly-bound state in the heavy-light subsystems alone. Moreover, we note that an interaction-independent behavior has been observed for all tested combinations of deeply-bound states in the expansion. It appears that not all deeply bound two-body states are equally important for the universal behavior of the three-body system.

Based on the analysis presented in this article we are convinced that the bound states in a three-body system near an excited-state two-body threshold display universality. However, in order to gain a better understanding of the universal behavior, additional insight into the mechanism of deeply bound two-body states would be desirable. Moreover, it would be interesting to analyze whether an odd-wave pseudopotential \cite{Girardeau2004,Sekino2021} can provide the universal limit for the three-body wave functions that we have observed for the odd-numbered thresholds. Such a pseudo-potential could further help in understanding the independence of the universal three-body energies on the symmetry of the weakly-bound heavy-light states. Then, it might be possible to establish a mapping between the results for even and odd-numbered thresholds. Additionally, the striking difference in the speed of convergence towards the universal limit between heavy bosons and fermions needs further analysis. Finally, the study of three-body resonances in this one-dimensional system poses an interesting and experimentally relevant task.

\acknowledgments
We are very grateful to P.~M.~A.~Mestrom and W.P.~Schleich for fruitful discussions. L.H. and M.A.E. thank the Center for Integrated Quantum Science and Technology (IQ$^{\rm ST}$) for financial support. The research of the IQ$^{\rm ST}$ is financially supported by the Ministry of Science, Research and Arts Baden-W\"urttemberg. The authors acknowledge support by the state of Baden-W\"urttemberg through bwHPC and the German Research Foundation (DFG) through grant no INST 40/467-1 FUGG (JUSTUS cluster) and INST 40/575-1 FUGG (JUSTUS 2 cluster).

\newpage
\appendix

\section{Methods}\label{app:formalism}
In this appendix we derive the integral equation, Eq.~\eqref{eq:FE_final_MainText}, corresponding to the three-body Schr\"odinger equation, Eq. \eqref{eq:3bodySGL} within the Faddeev approach \cite{Faddeev1961,Sitenko1991} and the separable expansion \cite{Weinberg1963,Sitenko1991,Mestrom2019}. We consider the three-body system as introduced in section \ref{sec:threebody1D}, where the light particle of mass $m$ is called particle 1, and particle 2 and 3 are the two identical heavy ones of mass $M$.

\subsection{The Faddeev equations}
We start by considering the three-body Schr\"odinger equation \eqref{eq:3bodySGL} in representation-free form
\begin{equation}\label{eq:3bodySGL_APP}
\left(H_0 + V_{31} + V_{12}\right) |\psi\rangle = \E |\psi\rangle.
\end{equation}
Here, $H_0$ is the kinetic energy operator without the center-of-mass motion, whereas $V_{31}$ and $V_{12}$ describe the pair-interactions between particles 3 and 1, and between particles 1 and 2, respectively.

According to the Faddeev approach \cite{Faddeev1961,Sitenko1991}, the solution of the Schr\"odinger equation \eqref{eq:3bodySGL_APP} is given by the superposition
\begin{equation}\label{eq:totalKetPsi}
|\psi\rangle \equiv |\phi^{(2)}\rangle + |\phi^{(3)}\rangle
\end{equation}
of $|\phi^{(2)}\rangle$ and $|\phi^{(3)}\rangle$, related to the so-called Faddeev components
\begin{subequations}\label{eq:FaddeevComponents}
\begin{align}
|\Phi^{(2)}\rangle &\equiv G_0^{-1} |\phi^{(2)}\rangle  \label{eq:FaddeevComponents_a} \\
|\Phi^{(3)}\rangle &\equiv G_0^{-1} |\phi^{(3)}\rangle \label{eq:FaddeevComponents_b}.
\end{align}
\end{subequations}
Here, ${G_0 = (\E-H_0)^{-1}}$ is the three-body Green function corresponding to $H_0$. The Faddeev components obey the Faddeev equations
\begin{subequations}\label{eq:FaddeevEQ}
\begin{align}
|\Phi^{(2)}\rangle &= T_{31} G_0\,|\Phi^{(3)}\rangle  \label{eq:FaddeevEQ_a}\\
|\Phi^{(3)}\rangle &= T_{12} G_0\,|\Phi^{(2)}\rangle,\label{eq:FaddeevEQ_b}
\end{align}
\end{subequations}
that is a system of two coupled integral equations.

The matrices $T_{31},\,T_{12}$ are the $T-$matrices of the three-body system if only the pair-interaction $V_{31}$ between particle $3$ and $1$, or $V_{12}$ between particle $1$ and $2$ respectively, is present. Hence, $T_{31}$ and $T_{12}$ fulfill the Lippmann-Schwinger equation \cite{Sitenko1991}
\begin{subequations}\label{eq:LST}
\begin{align}
T_{31} &= V_{31} + V_{31}G_0T_{31}  \label{eq:LST_a}\\
T_{12} &= V_{12} + V_{12}G_0T_{12}\label{eq:LST_b}.
\end{align}
\end{subequations}

\subsection{Momentum representation}\label{subsec:momentumrepresentation}
Three-body systems can be described in three different arrangements of relative motions, corresponding to three sets of so-called Jacobi momenta \cite{Sitenko1991}. For $\{i,j,l\} = \{1,2,3\}$ and the two cyclic permutations thereof we denote by
\begin{equation}
k_{ij} \equiv \frac{m_j k_i - m_i k_j}{m_i + m_j}
\end{equation}
the relative momentum between the particles $i$ and $j$, whereas 
\begin{equation}
p_l \equiv \frac{(m_i + m_j)k_{l} - m_l(k_i+k_j)}{m_i+m_j+m_l}
\end{equation}
is the momentum of particle $l$ relative to the center-of-mass of the pair of particles $i,j$. The momentum and mass of each particle are denoted by $k_i$ and $m_i$ respectively, with $i = 1,2,3$.

The relations between the Jacobi momenta read

%\begin{subequations}\label{eq:cs1}
\begin{align}
k_{12}(k_{23},p_1) &= -\frac{\alpha_y}{2}k_{23} + \alpha_x p_1 \nonumber \\
 p_3(k_{23},p_1) &= -k_{23} - \frac{1}{2}p_1 \label{eq:cs1a}%\\
\end{align}
%\end{subequations}

%\begin{subequations}\label{eq:cs1}
\begin{align}
k_{31}(k_{23},p_1) &= -\frac{\alpha_y}{2}k_{23} - \alpha_x p_1 \nonumber \\
 p_2(k_{23},p_1) &= k_{23} - \frac{1}{2}p_1 \label{eq:cs1b}
\end{align}
%\end{subequations}

%\begin{subequations}\label{eq:cs2}
\begin{align}
k_{12}(k_{31},p_2) &= -\frac{\alpha}{1+\alpha}k_{31} - \alpha_x \alpha_y p_2 \nonumber \\
 p_3(k_{31},p_2) &= k_{31} - \frac{\alpha}{1+\alpha}p_2, \label{eq:cs2a}
\end{align}
with the mass ratio $\alpha = M/m$ and the coefficients
\begin{equation}
\alpha_x \equiv \frac{1+2\alpha}{2(1+\alpha)}\qquad \mathrm{and}\qquad \alpha_y \equiv \frac{2}{1+\alpha}.
\end{equation}
%\end{subequations}

%\begin{subequations}\label{eq:cs2}
%\begin{align}
%k_{23}(k_{31},p_2) &= -\frac{1}{2}k_{31} + \frac{1+2\alpha}{2(1+\alpha)}p_2 \nonumber \\
%p_1(k_{31},p_2) &= -k_{31} - \frac{1}{1+\alpha}p_2 \label{eq:cs2b}
%\end{align}
%\end{subequations}

%\begin{subequations}\label{eq:cs3}
%\begin{align}
%k_{23} = -\frac{1}{2}k_{12} -& \frac{m+2M)}{2(m+M)}p_3 \nonumber \\
%& p_1 = k_{12} - \frac{m}{m+M}p_3 \label{eq:cs3a}  \\
%k_{31} = -\frac{M(m+M)}{(m+M)^2}k_{12} +& \frac{m(m+2M)}{(m+M)^2}p_3 \nonumber \\
%& p_2 = -k_{12} - \frac{M}{m+M}p_3. \label{eq:cs3b}
%\end{align}
%\end{subequations}

The system of Faddeev equations \eqref{eq:FaddeevEQ_a} and \eqref{eq:FaddeevEQ_b} can be reduced to a single equation by making use of the exchange symmetry between the two identical heavy particles in the three-body system. The exchange of particles 2 and 3 is most easily described in the set of Jacobi momenta $k_{23}$ and $p_1$ by the transformation $k_{23}\to-k_{23}$. From Eqs. \eqref{eq:cs1a} and \eqref{eq:cs1b} we deduce
\begin{subequations}
\begin{align}
k_{31}(-k_{23},p_1) &= -k_{12}(k_{23},p_1) \\
p_2(-k_{23},p_1) &= p_3(k_{23},p_1).
\end{align}
\end{subequations}

We distinguish two cases: the identical particles 2 and 3 are either bosons or fermions. Depending on this property, the total three-body wave function $\psi(-k_{23},p_1) = \pm \psi(k_{23},p_1)$ has to preserve (bosons) or flip (fermions) its sign under exchange of the two particles. This fact can be expressed in terms of the Faddeev components via Eq. \eqref{eq:totalKetPsi} which yields
\begin{align}
\Phi^{(2)}(-k_{12},p_3) + \Phi^{(3)}&(-k_{31},p_2)  = \nonumber \\
& \pm\left[\Phi^{(2)}(k_{31},p_2) + \Phi^{(3)}(k_{12},p_3)\right]
\end{align}
after the application of $G_0^{-1}$.

As a result, we can relate the two Faddeev components to each other
\begin{subequations}\label{eq:bosferm}
\begin{align}
\Phi^{(3)}(k_{12},p_3) &= \pm \Phi^{(2)}(-k_{12},p_3)\\
\Phi^{(3)}(k_{31},p_2) &= \pm \Phi^{(2)}(-k_{31},p_2).
\end{align}
\end{subequations}
With help of Eqs. \eqref{eq:cs1a} and \eqref{eq:cs1b} this allows us to cast the total wave function in the form
\begin{align}
\psi(k_{23},p_1) =~&\phi^{(2)}\left(-\frac{\alpha_y}{2}k_{23} - \alpha_x p_1,k_{23}-\frac{1}{2}p_1\right) \nonumber \\
\pm& \phi^{(2)}\left(\frac{\alpha_y}{2}k_{23} - \alpha_x p_1,-k_{23}-\frac{1}{2}p_1\right).
\end{align}

Since $\Phi^{(3)}$ can be obtained from $\Phi^{(2)}$ via Eq. \eqref{eq:bosferm}, it is sufficient to only obtain the Faddeev component $\Phi^{(2)}$ from its Faddeev equation \eqref{eq:FaddeevEQ_a}. In momentum representation this equation reads
\begin{align} \label{eq:integraleq}
\Phi^{(2)}&(k_{31},p_2) = \pm \iiiint \frac{\dint k_{31}' \dint p_2' \dint k_{31}'' \dint p_2''}{(2\pi)^4}  \nonumber \\
& \times \langle k_{31},p_2|T_{31}(\E)|k_{31}',p_2'\rangle \, \langle k_{31}',p_2'| G_0(\E) | k_{31}'',p_2''\rangle \nonumber \\
& \times \Phi^{(2)}\left[-k_{12}(k_{31}'',p_2''),p_3(k_{31}'',p_2'')\right].
\end{align}

Using the momentum representation of the free-particle three-body Green function \cite{Sitenko1991,Belyaev1990}
\begin{equation}\label{eq:greenfunction3}
\langle k',p'| G_0(\E) | k'',p''\rangle = (2\pi)^2\, \frac{\delta(k'-k'')\,\delta(p'-p'')}{\E-\frac{1}{2}k'^2 - \frac{1}{2}\alpha_x \alpha_y p'^2}
\end{equation}
as well as the relation of the three-body $T$-matrix to the off-shell two-body $t$-matrix \cite{Sitenko1991,Belyaev1990}
\begin{equation}
\langle k,p|T_{31}(\E)|k',p'\rangle =  2\pi\,\delta(p-p')\,t\left(k,k',\E-\frac{1}{2}\alpha_x \alpha_y p^2\right),
\end{equation}
allows us to perform the integrations over $p_2'$, $k_{31}''$ and $p_2''$ in Eq. \eqref{eq:integraleq}. Expressing further via Eq. \eqref{eq:cs2a} the momenta $-k_{12}$ and $p_3$ in terms of $k_{31}$ and $p_2$ for the arguments of $\Phi^{(2)}(-k_{12},p_3)$, we arrive at the integral equation
\begin{align} \label{eq:integraleqvarphi}
\Phi^{(2)}&(k,p) = \pm \int \frac{\dint k'}{2\pi} \frac{t\left(k,k',\E-\frac{1}{2}\alpha_x \alpha_y p^2\right)}{\E-\frac{1}{2}k'^2 - \frac{1}{2}\alpha_x \alpha_y p^2} \nonumber \\
& \times \Phi^{(2)}\left(\frac{\alpha}{1+\alpha} k' + \alpha_x \alpha_y p,k'-\frac{\alpha}{1+\alpha}p\right).
\end{align}
Here we have omitted the indices of the momenta $k_{31}$ and $p_2$.

By introducing the change of variables, $k' \equiv q + \frac{\alpha}{1+\alpha}p$, we cast Eq. \eqref{eq:integraleqvarphi} into the form
\begin{align}\label{eq:APPFE_momentum}
\Phi^{(2)}(k,p) = \pm \int \frac{\dint q}{2\pi}&\, \frac{t\left(k,q + \frac{\alpha}{1+\alpha}p,\E-\frac{1}{2}\alpha_x \alpha_y p^2\right)}{\E-\frac{1}{2} q^2 - \frac{1}{2}p^2 - \frac{\alpha}{1+\alpha}pq} \nonumber \\
&\times \Phi^{(2)}\left(p+\frac{\alpha}{1+\alpha}q,q\right)
\end{align}
eliminating the dependence on $p$ in the second argument of $\Phi^{(2)}$ inside the integral.

\ \newline
\subsection{Separable expansion}\label{app:separableexpansion}
We now transform Eq. \eqref{eq:APPFE_momentum} into the form of a Fredholm integral equation of the second kind. For most potentials, the off-shell $t$-matrix $t(k,k',\E)$ appearing in Eq.~\eqref{eq:APPFE_momentum} cannot be written as a product of functions of $k$ or $k'$ only. However, it is possible to expand the $t$-matrix into a sum of separable products \cite{Weinberg1963,Sitenko1991,Mestrom2019}
\begin{equation}\label{eq:APPsep_t}
t(k,k',\E) \equiv \sum_{\nu=0}^{\infty} \tau_\nu(\E)g_\nu(k,{\E})g_\nu(k',\E).
\end{equation}
Here, the functions $g_\nu$ and $\tau_\nu \equiv -\eta_\nu/(1-\eta_\nu)$ are determined by the integral equation
\begin{equation}\label{eq:APPetaeigenvalueeq}
\int \frac{\dint k'}{2\pi} V(k,k') \frac{1}{\E-\frac{1}{2}k'^2}\,g_\nu(k',\E) = \eta_\nu(\E) g_\nu(k,\E)
\end{equation}
with the momentum representation
\begin{equation}\label{eq:APPPotentialInMomentum}
V(k,k') = v_0 \int \dint \xi f(\xi) \e^{-\ii(k-k')\xi}
\end{equation}
of the heavy-light potential $v(\xi)= v_0 f(\xi)$. The functions $g_\nu$ are orthonormal with respect to the relation
\begin{equation}
\int \frac{\dint k}{2\pi} \frac{g_\nu(k,\E)\,g_{\nu'}(k,\E)}{\E-\frac{1}{2}k^2} = -\delta_{\nu,\nu'},
\end{equation}
where $\delta_{\nu,\nu'}$ denotes the Kronecker symbol.

By inserting Eq. \eqref{eq:APPsep_t} into Eq. \eqref{eq:APPFE_momentum}, we arrive at the expansion
\begin{equation}\label{eq:APPsep_phi}
\Phi^{(2)}(k,p) \equiv \pm \sum_{\nu=0}^{\infty} g_\nu(k,\E_p)\tau_\nu(\E_p)\varphi_\nu(p,\E)
\end{equation}
for $\Phi^{(2)}$ with $\E_p \equiv \E - \alpha_x \alpha_y p^2/2$ and
\begin{equation}\label{eq:phi_definition}
\varphi_\nu(p,\E) \equiv \int \dint q\, \frac{g_\nu\left(q+\frac{\alpha}{1+\alpha}p,\E_p\right) \Phi^{(2)}\left(p + \frac{\alpha}{1+\alpha}q,q\right)}{\E - \frac{1}{2} q^2 - \frac{1}{2}p^2 - \frac{\alpha}{1+\alpha}p q}.
\end{equation}

Finally, by substituting Eq. \eqref{eq:APPsep_phi} into Eq. \eqref{eq:phi_definition}, we derive a system of coupled integral equations
\begin{align}\label{eq:APPFE_final}
\varphi_\lambda(p,\E) = \pm &\sum_{\nu=0}^{\infty}\int \frac{\dint q}{2\pi}\, \tau_{\nu}(\E_{q})  \varphi_{\nu}(q,\E)  \nonumber \\
&\times \frac{g_{\lambda}\left(q+\frac{\alpha}{1+\alpha}p,\E_p\right)\, g_{\nu}\left(p + \frac{\alpha}{1+\alpha}q,\E_{q}\right)}{\E-\frac{1}{2}q^2 - \frac{1}{2}p^2 - \frac{\alpha}{1+\alpha}p q}
\end{align}
in only a single variable for the functions $\varphi_\nu(p,\E)$.

\bibliography{fewbody}

\end{document}